\documentclass[aps,prl,twocolumn, reprint,superscriptaddress]{revtex4-1}
\usepackage[utf8]{inputenx} 
\usepackage[english]{babel} 
\usepackage{hyperref} 
\usepackage{graphicx}
\usepackage{amsmath,amssymb}
\usepackage{dcolumn}
\usepackage{bm}
\usepackage{comment}
\usepackage{needspace}
\usepackage[export]{adjustbox}

\begin{document}

\title{Non-linearity in the system of quasiparticles of a superconducting resonator}
\author {M.~Vignati}
\email[Corresponding author: ]{marco.vignati@roma1.infn.it}
\affiliation{Sapienza Universit\`a di Roma -- Dipartimento di Fisica, I-00185, Roma, Italy}
\affiliation{Istituto Nazionale di Fisica Nucleare -- Sezione di Roma, I-00185, Roma, Italy}
\author{C.~Bellenghi}
\altaffiliation{Now at Technische Universit\"at M\"unchen, Physik-Department,  D-85748, Garching, Germany}
\affiliation{Sapienza Universit\`a di Roma -- Dipartimento di Fisica, I-00185, Roma, Italy}
\author{L.~Cardani}
\author{N.~Casali}
\affiliation{Istituto Nazionale di Fisica Nucleare -- Sezione di Roma, I-00185, Roma, Italy}
\author{I.~Colantoni}
\affiliation{Consiglio Nazionale delle Ricerche -- Istituto di Nanotecnologia, I-00185, Roma, Italy}
\affiliation{Istituto Nazionale di Fisica Nucleare -- Sezione di Roma, I-00185, Roma, Italy}
\author{A.~Cruciani}
\affiliation{Istituto Nazionale di Fisica Nucleare -- Sezione di Roma, I-00185, Roma, Italy}

\date{\today}
\begin{abstract}
We observed a strong non-linearity in the system of quasiparticles of a superconducting aluminum resonator,
due to the Cooper-pair breaking from the absorbed readout power. 
We observed both negative and positive feedback effects, controlled by the detuning of the readout frequency,
which are able to alter the relaxation time of quasiparticles by a factor greater than 10. 
We estimate that  the $(70\pm5)$~\% of the total non-linearity of the device  is due to quasiparticles. 
\end{abstract}
\maketitle

Superconducting resonators are used to build sensitive detectors, amplifiers and quantum circuits. 
These devices base their working principle on
non-linear inductances, which can be engineered via Josephson Junctions~\cite{DevoretQ,paramp,stj} 
or can rely on the intrinsic kinetic inductance of the superconductor~\cite{Day:2003fk,Ho-Eom:2012nr,kineticon}.

 In these applications, the superconductor is cooled well below its critical temperature,
 almost all electrons are bound in Cooper pairs, and the circuit is in principle lossless.
Photon or phonon interactions, however, can break the pairs and create quasiparticles, 
which then recombine on timescales that decrease with their density~\cite{kaplan}. 
The presence of quasiparticles, along with two-level systems (TLS)~\cite{MartinisTLS,mcrae}, is one of the main source of losses and can limit the quality factor of the resonator.

The readout power absorbed in the circuit can also break the pairs and increase the density of quasiparticles, in a way similar to a temperature increase~\cite{visser2014}. This leads to the establishment of an electro-thermal feedback due to the variation of the absorbed power with the density of quasiparticles~\cite{visser2010,Thompson_2013,Thomas_2015}.  In this work we report on the first observation of a non-linear behavior due to the electro-thermal feedback and its effect on the relaxation time of quasiparticles.
\\
\needspace{0.5\baselineskip}
\noindent
The resonator under study consists of a lumped-element LC circuit coupled to a coplanar wave guide, realized with a 60 nm aluminum lithography 
on a 2x2~cm$^2$ wide, 0.3 mm thick silicon substrate. 
The  inductor L is 6~cm long and 60~$\mu$m wide, and it is winded in a meander with 5~$\mu$m spacing and closed on a 2-finger capacitor C at distance of 60~$\mu$m.
The resonator is analogous to that presented in Ref.~\cite{Cardani:2017qr}, and is operated as phonon-mediated~\cite{swenson} kinetic inductance detector~\cite{Day:2003fk} in a cryogen-free dilution refrigerator with base temperature of 20 mK. The transmission $S_{21}$ of the circuit is measured by means of a heterodyne electronics~\cite{Bourrion:2013ifa}. The low-power, low-temperature resonant frequency and quality factor are found to be $f_{r,0}=2.556363~{\rm GHz}$ and $Q_0=98100$, respectively, and the fraction of kinetic over total inductance is $\alpha = 2.5$~\%.  

\begin{figure}[t]
\begin{center}
\includegraphics[width=1\linewidth, left]{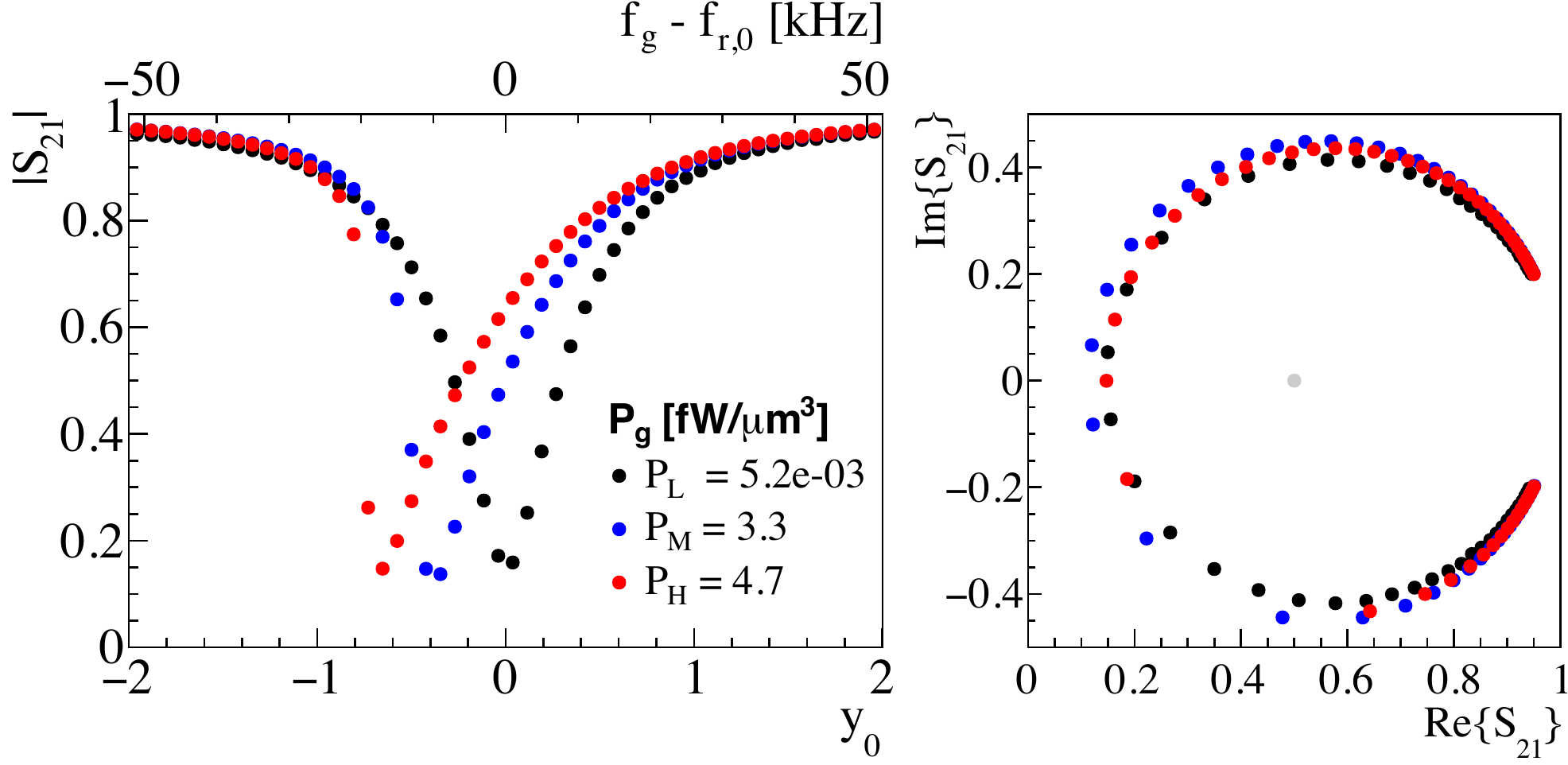} 
\includegraphics[width=1\linewidth, left]{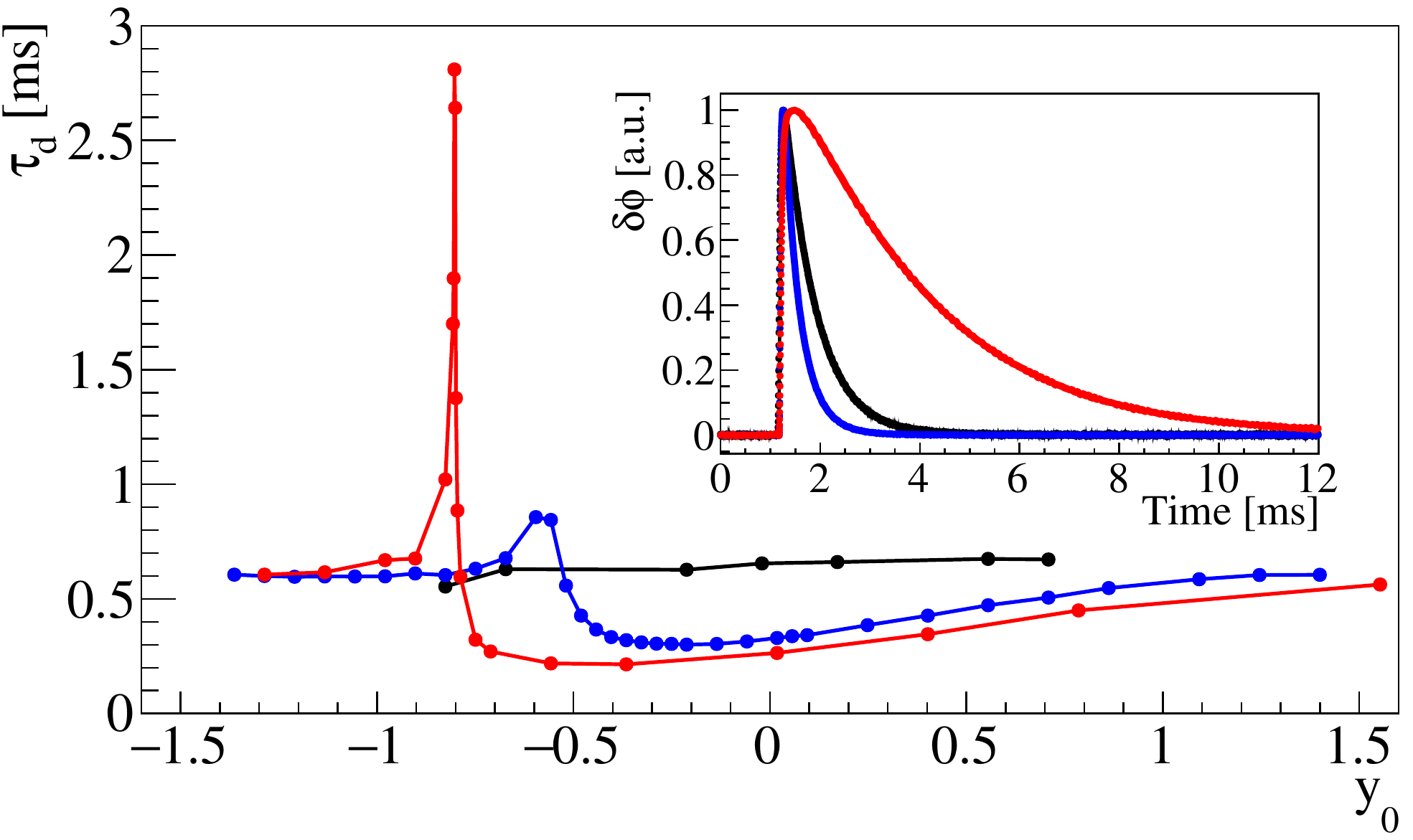} 
\caption{Frequency scan of the resonator at different generator powers $P_g$. Magnitude of the transmission $|S_{21}|$  as a function
of the  detuning in line-widths ($y_0$)  (top, left), real  and imaginary part of $S_{21}$ (top, right), and decay time of the pulses $\tau_{d}$ as a function of $y_0$ (bottom); 
Inset:  $\delta\phi$ average pulses at low and medium powers with $y_0 \sim 0$, and at high power with $y_0$ corresponding to the maximum of the $\tau_d$ overshoot.}
\label{fig:fig1}
\end{center}
\end{figure}

Sweeps of the generator frequency $f_g$ across $f_{r,0}$ at three power levels, here quoted in terms of density over the inductor volume  ($P_g=5.2$~aW/$\mu m^3$, 3.3 and 4.7 fW/$\mu m^3$), are shown in Fig.~\ref{fig:fig1} (top).
In the left panel of the figure the magnitude of the transmission $S_{21}$ past the chip, corrected for line effects~\cite{khalil}, 
is shown as a function of the detuning in line-widths units, $y_0 = Q_0 (f_g - f_{r,0})/f_{r,0}$, while the right panel shows the real and imaginary parts of $S_{21}$. 
The resonance moves significantly to lower frequencies and becomes asymmetric with increasing power, a known effect due to the  kinetic inductance non-linearity. 
At even higher powers, not shown in the figure, the resonance enters in a hysteretic regime, implying that a fraction of the transmission is no longer accessible~\cite{visser2010, swenson2013}.  The size of the circle in the real and imaginary plane of $S_{21}$, which is directly proportional to the quality factor of the resonator, changes less evidently. It increases at medium power, as expected with the saturation of TLS, and decreases back at high power, an effect that is attributed to the increase of the number of quasiparticles. 

In order to excite the resonator, the light from a room-temperature pulsed LED is driven to one side of the silicon substrate by an optical fiber passing through the cryostat. The  absorbed light is converted to phonons in the silicon, which scatter through the lattice until they are absorbed in the superconductor and break the Cooper pairs. 
The pair-breaking alters the resonator frequency and quality factor, which are in turn measured through the phase  ($\delta\phi$)  and amplitude ($\delta A $)  variations of the wave transmitted past the resonator relative to the center of the $S_{21}$ circle (see e.g.~\cite{zmu_annrev2012}). 
Pulses following the optical excitation are acquired with a software trigger and averaged over around 500 samples to reduce noise in the estimation of their shape.
The rise time of the pulses is dominated by the ring time of the resonator  ($Q/\pi f_r = 12~\mu s$) and  by the phonon life-time in the substrate ($\sim 15~\mu s$), while the decay time is attributed to the relaxation of quasiparticles~\cite{martinez2019}. The duration of the LED excitation is  $2~\mu$s, and does not contribute significantly to the shape of the pulses.  

The decay time $\tau_d$ of the pulses, measured as the time difference between the  90\% and the 30\% of their trailing edge, is shown in the bottom panel of Fig.~\ref{fig:fig1} as a function of the detuning $y_0$. At low power $\tau_d$ does not depend on $y_0$ and averages to $0.64$~ms. The dependency is instead sizeable at medium and high powers, and in the latter case  it varies from a minimum of $0.21$ to a maximum of $2.8$~ms. The overshoot at high power is found to be very narrow,  around $0.006$ line-widths corresponding to 160 Hz in our case, and is therefore unlikely to spot if not intentionally searched for. 
\\
\needspace{0.5\baselineskip}
\noindent
In the following we present a model which guided us in the measurement of the presented data 
and which is used in this work to identify origin and  amount of the non-linearity. 
We begin with the study of the frequency sweeps and then move to the decay time of the pulses.
 
When $N_{qp}$ quasiparticles are created via pair-breaking, before recombining they store an amount of energy
\begin{equation}
E_{qp} = N_{qp}\Delta 
\end{equation}
where $2\Delta$ is the binding energy of a Cooper pair.  In superconducting resonators the presence of quasiparticles, 
to a first-order approximation, modifies resonant frequency $f_r$ and quality factor $Q$  with respect to their low-power and low-temperature values ($f_{r,0},Q_0$) as:
\begin{eqnarray}
 x_r  &=& -  \alpha \frac{E_{qp}}{E'} \label{eq:res}\\
\frac{1}{Q} &=& \frac{1}{Q_0} +   2\frac{\alpha}{\beta}  \frac{E_{qp}}{E'}   \label{eq:qf}
\end{eqnarray}
where $x_r = ({f_r - f_{r,0}})/{f_{r,0}}$,  $\beta/2$ is the ratio between the frequency and inverse quality factor responses and $E'$ is expected
to be of the order of the pairing energy of the superconductor.
The   measurable quantity  is the transmission $S_{21}$ which, for the resonator under study, can be expressed as~\cite{zmu_annrev2012}:
\begin{eqnarray}\label{eq:S21}
S_{21} =1 - \frac{Q}{Q_c}\frac{1}{1+2 j Qx}
\end{eqnarray}
where  $Q_c$ is the coupling quality factor and $x = (f_g - f_r) / f_r$  is the detuning of the generator frequency with respect to $f_r$. 
In turn $x$ can be expressed in terms of the detuning $x_0$ with respect to $f_{r,0}$, $x_0 = (f_g - f_{r,0})/f_{r,0}$,  as:
\begin{equation}\label{eq:det}
x \simeq x_0 +  \alpha\frac{E_{qp}}{E'} 
\end{equation}
where in the calculation we approximated $1+x_r\simeq1$. 

The energy absorbed in quasiparticles from the readout power amounts to~\cite{zmu_annrev2012,visserwit}:
\begin{eqnarray}
E_{qp}^P &=& \eta_{g} P_{qp} \tau_{qp}\label{eq:eqp}\\
P_{qp} &=&  P_{g} \frac{2 Q^2}{Q_c Q_{qp}} \frac{1}{1+4 Q^2 x^2}\label{eq:pqp}
\end{eqnarray}
where $\eta_{g}$ is the efficiency in the creation of  quasiparticles~\cite{deVisserEta}
\footnote{$\eta_g$ depends on the number of quasiparticles, and thus  on $P_{qp}$~\protect\cite{Goldie2013}. This dependency can be neglected in our first-order model.}, 
$\tau_{qp}$ and $Q_{qp}$ are the recombination time and the internal quality factor of quasiparticles, respectively.
We finally define the non-linearity parameter as:
\begin{equation}\label{eq:aqp}
a_{qp} =  \alpha Q \frac{E_{qp}^P(x=0)}{E'}
\end{equation}

The energy $E_{qp}^P$, may represent only a fraction of the total absorbed energy $E^P$ and thus of the non-linearity.
In our device the total non-linearity ($a$) manifests itself in the non-linearity of the kinetic inductance, which has been already extensively studied~\cite{swenson,zmu_annrev2012}:
\begin{equation}\label{eq:atot}
a = 
\alpha Q \frac{E^P}{E'} = 
\alpha \frac{2 Q^3}{Q_c} \frac{P_{g}}{2\pi f_{r,0}}\frac{\gamma}{E'}
\end{equation}
where $\gamma$ is dimensionless  parameter of order 1 to account for a possible difference in the energy scale.
Defining $y=Qx$, we can rewrite Eqns.~\ref{eq:det} and \ref{eq:qf} in terms of $a$ and $a_{qp}$ as:
\begin{eqnarray}
y &=& y_0 z +  \frac{a_{0} z^3}{1+4y^2}\label{eq:ytot}\\
z &=& \frac{Q}{Q_0} = 1-   \frac{2}{\beta}\frac{a_{qp,0}z^3}{1+4y^2}\label{eq:z}
\end{eqnarray}
where $a_{0}=a(Q=Q_0)$ and $a_{qp,0}=a_{qp}(Q=Q_0)$.
These equations describe the detuning in line-widths with respect to the power-shifted resonant frequency ($y$),
and the fractional change of the quality factor ($z$) as a function of $y$, respectively.  It has to be underlined that the detuning is affected by the  total non-linearity ($a$),
while the quality factor is affected only by the portion of non-linearity due to quasiparticles ($a_{qp}$).

\begin{figure}[t]
\begin{center}
\includegraphics[width=1\linewidth]{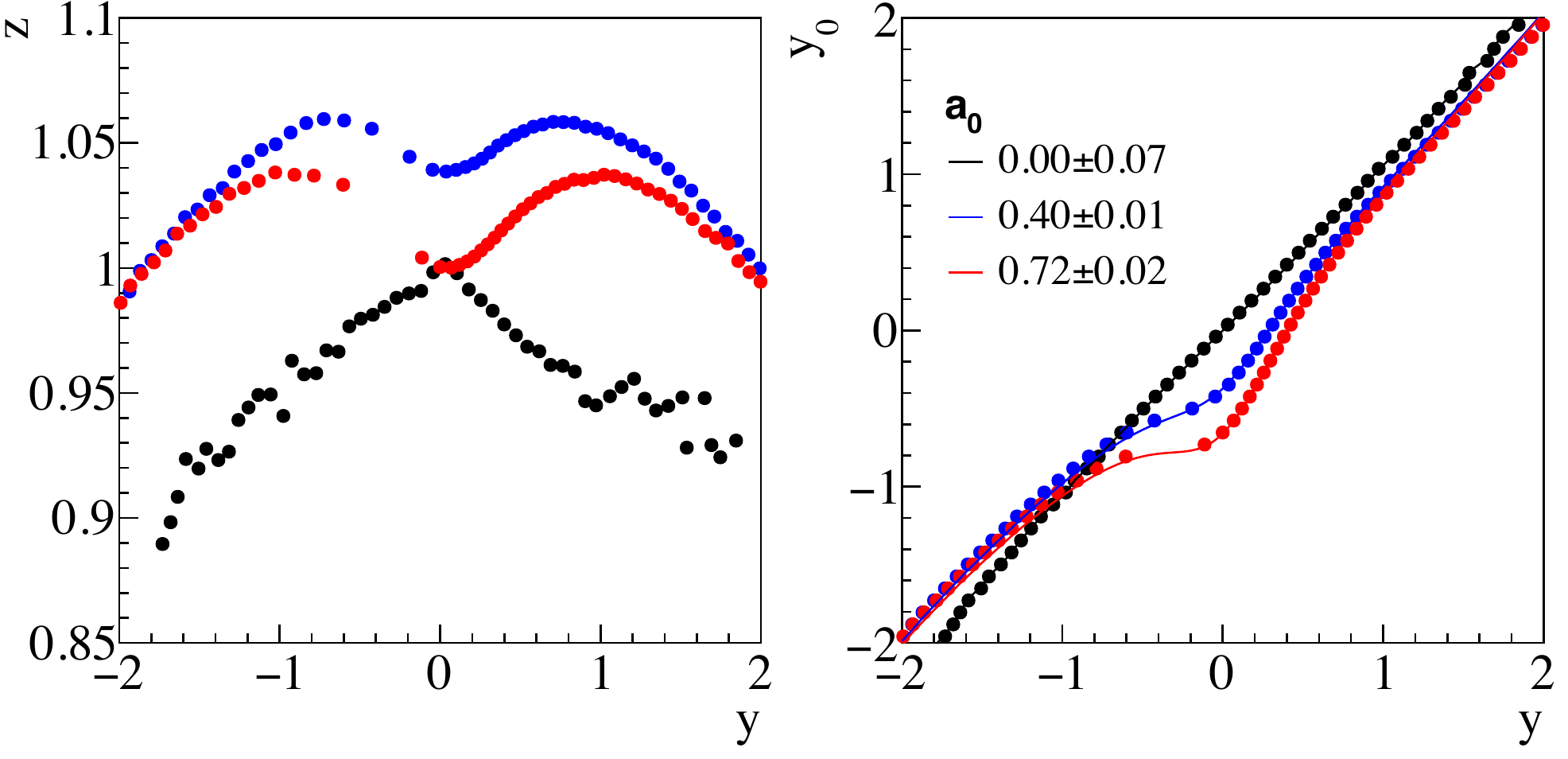} 
\caption{Left:  $z=Q/Q_{0}$ as a function of the detuning with respect to the power-shifted resonant frequency ($y$) at low (black), medium (blue) and high (red) powers; Right: $y_0$ as a function of $y$ (dots) along with fits for $a_0$ of Eq.~\ref{eq:ytot}  (lines).}
\label{fig:fig2}
\end{center}
\end{figure}

The values of ($z,y$)  in Eq.~\ref{eq:ytot}  are calculated from the $S_{21}$ data (cf. Suppl. Mat.) and  shown in  Fig.~\ref{fig:fig2} (left).
 At low-power the value of $z$ decreases with $|y|$, which reveals that the quality factor is dominated by TLS as it lowers with decreasing absorbed power. At higher powers the same behavior is observed at high $|y|$, 
 while for $|y|$ around zero it decreases because of the power-generated quasiparticles. It has to be noticed that, at least for the resonator under study, the $z$ variation with $y$ at fixed power is only at few \% level, presumably because of a compensation between the quality factors of quasiparticles and TLS.
  If quasiparticles had dominated the quality factor, we could have used Eq.~\ref{eq:z} to extract $a_{qp,0}$ from fits to the $(z,y)$ data, provided that $\beta$ is measured independently. Nevertheless, as it will be shown next, it is possible to estimate $a_{qp,0}$ directly from the decay-time of the pulses, without introducing the effect of TLS in the model, and thus other free parameters.

The  points in the ($y,y_0$) plane  are shown in Fig.~\ref{fig:fig2} (right) along with fits of Eq.~\ref{eq:ytot} for $a_0$ and using values of $z$ from the $(z,y)$ graph. The results are $a_0 =0.00\pm0.07, 0.40\pm0.01$ and $0.72\pm0.02$ for the low, medium and high powers, respectively,  with a 6\% systematic error  added from the model (cf. Suppl. Mat.).  
\\
\needspace{0.5\baselineskip}
\noindent
The presence of a population of quasiparticles at equilibrium, along the frequency and quality factor of the resonator in Eqns.~\ref{eq:res} and \ref{eq:qf}, modifies the recombination time as $\tau_{qp} \propto 1/ N_{qp}$~\cite{kaplan}.  One can therefore map the recombination time and the shift of the resonant frequency as: 
\begin{equation}\label{eq:tauqp}
\frac{1}{\tau_{qp}} = \frac{1}{\tau_{qp,0}} - \frac{y_r}{\tau_k}
\end{equation}
where $\tau_k$ embeds the physics governing the dependency of frequency and recombination time with $N_{qp}$,  $\tau_{qp,0}$ is the saturation value of the recombination time at low-power and low-temperature~\cite{BarendsTau}, and
\begin{equation}\label{eq:yr}
y_r=- \frac{a_{qp,0} z^3}{1+4y^2}
\end{equation}
is the shift of the resonant frequency due to the power absorbed in quasiparticles.

When there are $M$ quasiparticles out of equilibrium, as in the case of excitation with the light pulses, their time evolution can be described as
\begin{equation}\label{eq:dmdt}
\frac{dM}{dt} = -\frac{M}{\tau_{qp}}\left(1-\frac{ \delta N^P_{qp}}{M}\right) 
\end{equation}
where the first term of the product accounts for the recombination, while the second for the extra quasiparticles $\delta N^P_{qp}$ injected or removed by the variation of the absorbed power.
With a first order approximation~\footnote{In the calculation of derivative, we approximated $\frac{d(\tau_{qp}/Q_{qp})}{dM}=0$ since the dependency of $\tau_{qp}$ and $Q_{qp}$ on the number of quasiparticles cancels to a good approximation~\protect\cite{deVisserEta}}
we obtain from Eq.~\ref{eq:eqp}:
\begin{eqnarray}
\frac{ \delta N^P_{qp}}{M} &=& \frac{\delta E^P_{qp}}{M\Delta }=-\frac{\delta x}{\delta x_{qp}} a_{qp}{F(y)}\label{eq:nqpdm}\\
F(y) &=& \frac{8y}{(1+4y^2)^2} + \frac{\delta Q^{-1}}{\delta x} \frac{2}{(1+4y^2)^2}\label{eq:feedback}
\end{eqnarray}
where $\delta x$ and $\delta Q^{-1}$ are the variations in detuning and inverse quality factor following the excitation, respectively, $\delta x_{qp} = \alpha M\Delta/ E'$ is the contribution to  $\delta x$ due only to quasiparticles, $F(y)$ accounts for the feedback sign and dependency on $y$, and $a_{qp}$ accounts for its intensity. 

Putting together Eqns.~\ref{eq:dmdt} and \ref{eq:nqpdm} we obtain the expression for the relaxation time:
\begin{equation}\label{eq:efftau2}
\tau_{rel} = \frac{\tau_{qp}}{1-\frac{ \delta N^P_{qp}}{M}} 
= \frac{\tau_{qp}} {1 +\frac{\delta x}{\delta x_{qp}}a_{qp} F(y)}
\end{equation}
In presence of quasiparticles' non-linearity only, $\delta x/\delta x_{qp}=1$ and $\delta Q^{-1}/\delta x=2/\beta$. Other non-linearities, however, alter these parameters by adding a dependency on $y$.  Including these effects would add free parameters to the model, therefore we chose to estimate $\delta x$ and $\delta Q^{-1}/\delta x$ directly from the   $\delta \phi$ and $\delta A$ pulses acquired at each $y$ (cf. Suppl. Mat.).  The value of $\delta x_{qp}$ is estimated  from the pulses at high $|y|$, where the non-linearities are suppressed. 

\begin{figure}[t]
\begin{center}
\includegraphics[width=1\linewidth, left]{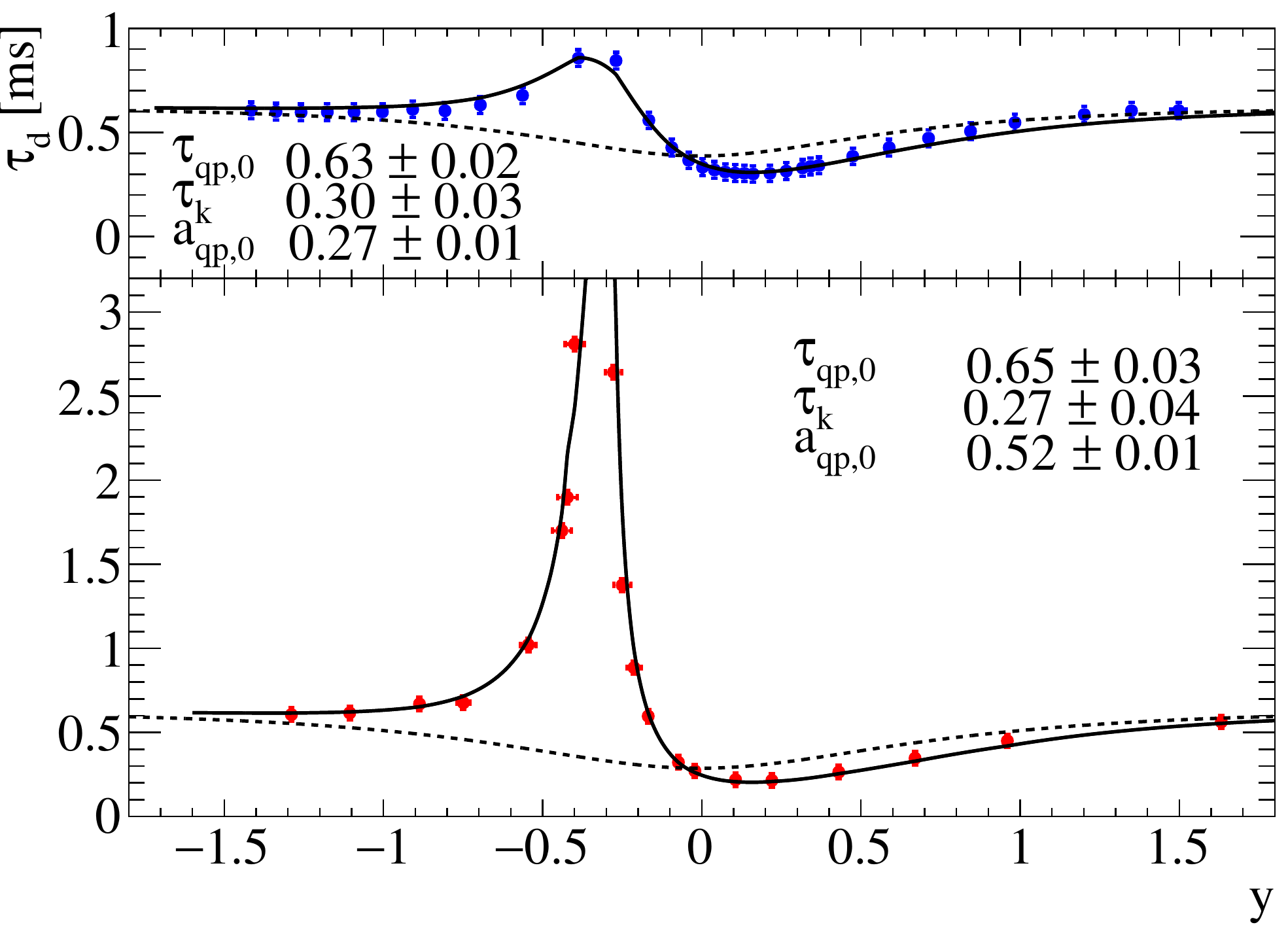} 
\caption{Decay time of the pulses ($\tau_{d}$) as a function of the power-shifted detuning ($y$) for the medium (top) and high (bottom) powers. The solid lines are fits of the model for the relaxation time including feedback effects (Eq.~\ref{eq:efftau2}). The dashed lines indicate the contribution to these fits due to the steady change of $\tau_{qp}$ with absorbed power (Eq.~\ref{eq:tauqp}) . }
\label{fig:fig3}
\end{center}
\end{figure}

We fit the measured decay time of signals as a function of $y$ with $\tau_{qp,0}, \tau_k$ and $a_{qp,0}$ as free parameters (Fig.~\ref{fig:fig3}). From the figure one can see that the first-order model we proposed reproduces well the data with $a_{qp,0}=0.27\pm0.01$ and $a_{qp,0}=0.52\pm0.01$ for the medium and high powers, respectively. We note that $a_{qp}/a= a_{qp,0}/a_{0}=(68\pm3)~\%$ and $(72\pm3)~\%$ for the
 the medium and high powers, respectively, pointing to the fact that quasiparticles account for a large fraction of the total non-linearity.  
 Combining the two measurements and including the systematic error from the model we obtain $a_{qp}/a=(70\pm5)~\%$.

\begin{figure}[t]
\begin{center}
\includegraphics[width=1\linewidth, left]{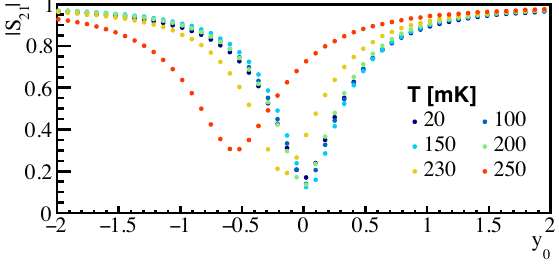} 
\includegraphics[width=1\linewidth, left]{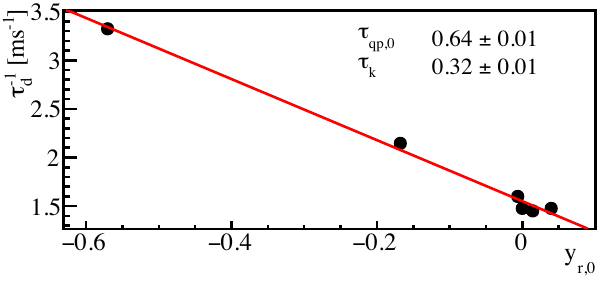} 
\caption{Temperature and frequency scan of the resonator at low generator power. Magnitude of the transmission $|S_{21}|$  as a function
of $y_0$ (top) and inverse of the decay time of the signals $\tau_{d}$  as a function of the shift of the resonant frequency in line-widths ($y_{r,0}$)  at each temperature (bottom). The solid line is a fit for Eq.~\ref{eq:tauqp}.}
\label{fig:fig4}
\end{center}
\end{figure}

In order to deepen the understanding of the observed phenomena,  we also study the variation of the decay time with temperature when the resonator is biased at low power. 
By doing this, we  remove the generator feedback and,  isolate the behaviour of $\tau_{qp}$ with a steady population of quasiparticles. 
Figure~\ref{fig:fig4} shows the resonance shift with temperature (top) and the decay time as a function of the resonant frequency shift (bottom).
Fitting Eq.~\ref{eq:tauqp} to the data points we find  $\tau_{qp,0}=0.64\pm 0.01$ and $\tau_k = 0.32 \pm 0.01$ ms, which are in  agreement with the values found from the
power-generated quasiparticles in Fig~\ref{fig:fig3}.

The ratio of  quasiparticles' (Eq.~\ref{eq:aqp}) to total  (Eq.~\ref{eq:atot}) non-linearities,
\begin{equation}\label{eq:aratio}
\frac{a_{qp}}{a} = \frac{\eta_{g}}{\gamma} \frac{2\pi f_{r,0}\tau_{qp}}{Q_{qp}}\,, 
\end{equation}
allows in principle to derive $\eta_{g}/\gamma$, provided that $Q_{qp}$ is known at least to some approximation.  
From the definition of the total quality factor, $Q^{-1} = Q_c^{-1} + Q_i^{-1} $, and assuming that the internal quality factor, $Q_i$, is dominated by TLS and quasiparticles,
${Q_i}^{-1} \simeq {Q_{TLS}}^{-1} + Q_{qp}^{-1}$,
we can argue that the maxima of $z$ in the frequency sweeps of Fig.~\ref{fig:fig2} (left) correspond to $Q_{TLS}\simeq Q_{qp}$. The value of $Q_{qp}$ can be therefore
estimated as:
\begin{equation}
\frac{1}{Q_{qp}}\Bigr|_{z_{\rm max}} \simeq \frac{1}{2}\left(\frac{1}{z_{\rm max}Q_0}-\frac{1}{Q_c}\right)
\end{equation} 
Using the values of $\tau_{qp}$ calculated with Eqns.~\ref{eq:tauqp} and ~\ref{eq:yr} in the point  $(y|z_{max}, z_{max})$ and the measured value of $Q_c$ ($Q_c = 115100$, cf. Suppl .Mat.), from Eq.~\ref{eq:aratio} we obtain $\eta_g/\gamma =(18\pm3)~\%$ and $(16\pm3)$~\% for the medium
and high power data, respectively. Assuming $\gamma = 1$,  the value of $\eta_g$ is in line with the predictions in Ref.~\cite{Goldie2013}.
\\
\needspace{0.5\baselineskip}
\noindent
Our results  reveal  the existence of a population of quasiparticles generated from the readout power which undergoes a strong electro-thermal feedback and which significantly modifies the properties of the superconducting circuit. As an example, the positive feedback could be exploited to increase the response of superconducting circuits at signal frequencies below $1/2\pi\tau_{rel}$
. The negative feedback, instead, could be exploited to make the circuit more resistant to quasiparticles' perturbations.  Materials with different intrinsic values
of $\tau_{qp}/Q_{qp}$ could be studied to enhance or suppress the non-linearity from quasiparticles.
\\
\needspace{0.5\baselineskip}
\noindent
The authors acknowledge useful discussion with J. Lorenzana, A. Monfardini and  I. M. Pop,
and thank the personnel of INFN Sezione di Roma for the technical support, in particular A. Girardi and M. Iannone.
This work was supported by the European Research Council (FP7/2007-2013) under Contract No. CALDER 335359 and by the Italian Ministry of Research under the FIRB Contract No. RBFR1269SL. 

\bibliographystyle{apsrev4-1}
\bibliography{../../calder}

\begin{thebibliography}{26}%
\makeatletter
\providecommand \@ifxundefined [1]{%
 \@ifx{#1\undefined}
}%
\providecommand \@ifnum [1]{%
 \ifnum #1\expandafter \@firstoftwo
 \else \expandafter \@secondoftwo
 \fi
}%
\providecommand \@ifx [1]{%
 \ifx #1\expandafter \@firstoftwo
 \else \expandafter \@secondoftwo
 \fi
}%
\providecommand \natexlab [1]{#1}%
\providecommand \enquote  [1]{``#1''}%
\providecommand \bibnamefont  [1]{#1}%
\providecommand \bibfnamefont [1]{#1}%
\providecommand \citenamefont [1]{#1}%
\providecommand \href@noop [0]{\@secondoftwo}%
\providecommand \href [0]{\begingroup \@sanitize@url \@href}%
\providecommand \@href[1]{\@@startlink{#1}\@@href}%
\providecommand \@@href[1]{\endgroup#1\@@endlink}%
\providecommand \@sanitize@url [0]{\catcode `\\12\catcode `\$12\catcode
  `\&12\catcode `\#12\catcode `\^12\catcode `\_12\catcode `\%12\relax}%
\providecommand \@@startlink[1]{}%
\providecommand \@@endlink[0]{}%
\providecommand \url  [0]{\begingroup\@sanitize@url \@url }%
\providecommand \@url [1]{\endgroup\@href {#1}{\urlprefix }}%
\providecommand \urlprefix  [0]{URL }%
\providecommand \Eprint [0]{\href }%
\@ifxundefined \urlstyle {%
  \providecommand \doi  [0]{\begingroup \@sanitize@url \@doi}%
  \providecommand \@doi [1]{\endgroup \@@startlink {\doibase
  #1}doi:\discretionary {}{}{}#1\@@endlink }%
}{%
  \providecommand \doi  [0]{doi:\discretionary{}{}{}\begingroup
  \urlstyle{rm}\Url }%
}%
\providecommand \doibase [0]{http://dx.doi.org/}%
\providecommand \Doi [0]{\begingroup \@sanitize@url \@Doi }%
\providecommand \@Doi  [1]{\endgroup\@@startlink{\doibase#1}\@@Doi}%
\providecommand \@@Doi [1]{#1\@@endlink}%
\providecommand \selectlanguage [0]{\@gobble}%
\providecommand \bibinfo  [0]{\@secondoftwo}%
\providecommand \bibfield  [0]{\@secondoftwo}%
\providecommand \translation [1]{[#1]}%
\providecommand \BibitemOpen [0]{}%
\providecommand \bibitemStop [0]{}%
\providecommand \bibitemNoStop [0]{.\EOS\space}%
\providecommand \EOS [0]{\spacefactor3000\relax}%
\providecommand \BibitemShut  [1]{\csname bibitem#1\endcsname}%
\bibitem [{\citenamefont {Martinis}\ \emph {et~al.}(1985)\citenamefont
  {Martinis}, \citenamefont {Devoret},\ and\ \citenamefont
  {Clarke}}]{DevoretQ}%
  \BibitemOpen
  \bibfield  {author} {\bibinfo {author} {\bibfnamefont {J.~M.}\ \bibnamefont
  {Martinis}}, \bibinfo {author} {\bibfnamefont {M.~H.}\ \bibnamefont
  {Devoret}}, \ and\ \bibinfo {author} {\bibfnamefont {J.}~\bibnamefont
  {Clarke}},\ }\Doi {10.1103/PhysRevLett.55.1543} {\bibfield  {journal}
  {\bibinfo  {journal} {Phys. Rev. Lett.},\ }\textbf {\bibinfo {volume} {55}},\
  \bibinfo {pages} {1543} (\bibinfo {year} {1985})}\BibitemShut {NoStop}%
\bibitem [{\citenamefont {Feldman}\ \emph {et~al.}(1975)\citenamefont
  {Feldman}, \citenamefont {Parrish},\ and\ \citenamefont {Chiao}}]{paramp}%
  \BibitemOpen
  \bibfield  {author} {\bibinfo {author} {\bibfnamefont {M.~J.}\ \bibnamefont
  {Feldman}}, \bibinfo {author} {\bibfnamefont {P.~T.}\ \bibnamefont
  {Parrish}}, \ and\ \bibinfo {author} {\bibfnamefont {R.~Y.}\ \bibnamefont
  {Chiao}},\ }\Doi {10.1063/1.322157} {\bibfield  {journal} {\bibinfo
  {journal} {Journal of Applied Physics},\ }\textbf {\bibinfo {volume} {46}},\
  \bibinfo {pages} {4031} (\bibinfo {year} {1975})}\BibitemShut {NoStop}%
\bibitem [{\citenamefont {Dolan}\ \emph {et~al.}(1979)\citenamefont {Dolan},
  \citenamefont {Phillips},\ and\ \citenamefont {Woody}}]{stj}%
  \BibitemOpen
  \bibfield  {author} {\bibinfo {author} {\bibfnamefont {G.~J.}\ \bibnamefont
  {Dolan}}, \bibinfo {author} {\bibfnamefont {T.~G.}\ \bibnamefont {Phillips}},
  \ and\ \bibinfo {author} {\bibfnamefont {D.~P.}\ \bibnamefont {Woody}},\
  }\Doi {10.1063/1.90783} {\bibfield  {journal} {\bibinfo  {journal} {Applied
  Physics Letters},\ }\textbf {\bibinfo {volume} {34}},\ \bibinfo {pages} {347}
  (\bibinfo {year} {1979})}\BibitemShut {NoStop}%
\bibitem [{\citenamefont {Day}\ \emph {et~al.}(2003)\citenamefont {Day},
  \citenamefont {LeDuc}, \citenamefont {Mazin}, \citenamefont {Vayonakis},\
  and\ \citenamefont {Zmuidzinas}}]{Day:2003fk}%
  \BibitemOpen
  \bibfield  {author} {\bibinfo {author} {\bibfnamefont {P.~K.}\ \bibnamefont
  {Day}}, \bibinfo {author} {\bibfnamefont {H.~G.}\ \bibnamefont {LeDuc}},
  \bibinfo {author} {\bibfnamefont {B.~A.}\ \bibnamefont {Mazin}}, \bibinfo
  {author} {\bibfnamefont {A.}~\bibnamefont {Vayonakis}}, \ and\ \bibinfo
  {author} {\bibfnamefont {J.}~\bibnamefont {Zmuidzinas}},\ }\Doi
  {10.1038/nature02037} {\bibfield  {journal} {\bibinfo  {journal} {Nature},\
  }\textbf {\bibinfo {volume} {425}},\ \bibinfo {pages} {817} (\bibinfo {year}
  {2003})}\BibitemShut {NoStop}%
\bibitem [{\citenamefont {Ho~Eom}\ \emph {et~al.}(2012)\citenamefont {Ho~Eom},
  \citenamefont {Day}, \citenamefont {LeDuc},\ and\ \citenamefont
  {Zmuidzinas}}]{Ho-Eom:2012nr}%
  \BibitemOpen
  \bibfield  {author} {\bibinfo {author} {\bibfnamefont {B.}~\bibnamefont
  {Ho~Eom}}, \bibinfo {author} {\bibfnamefont {P.~K.}\ \bibnamefont {Day}},
  \bibinfo {author} {\bibfnamefont {H.~G.}\ \bibnamefont {LeDuc}}, \ and\
  \bibinfo {author} {\bibfnamefont {J.}~\bibnamefont {Zmuidzinas}},\ }\Doi
  {10.1038/nphys2356} {\bibfield  {journal} {\bibinfo  {journal} {Nature
  Physics},\ }\textbf {\bibinfo {volume} {8}},\ \bibinfo {pages} {623}
  (\bibinfo {year} {2012})}\BibitemShut {NoStop}%
\bibitem [{\citenamefont {Faramarzi}\ \emph {et~al.}(2021)\citenamefont
  {Faramarzi} \emph {et~al.}}]{kineticon}%
  \BibitemOpen
  \bibfield  {author} {\bibinfo {author} {\bibfnamefont {F.~B.}\ \bibnamefont
  {Faramarzi}} \emph {et~al.},\ }\href@noop {} { (\bibinfo {year} {2021})},\
  \Eprint {http://arxiv.org/abs/2012.08654} {arXiv:2012.08654} \BibitemShut
  {NoStop}%
\bibitem [{\citenamefont {Kaplan}\ \emph {et~al.}(1976)\citenamefont {Kaplan},
  \citenamefont {Chi}, \citenamefont {Langenberg}, \citenamefont {Chang},
  \citenamefont {Jafarey},\ and\ \citenamefont {Scalapino}}]{kaplan}%
  \BibitemOpen
  \bibfield  {author} {\bibinfo {author} {\bibfnamefont {S.~B.}\ \bibnamefont
  {Kaplan}}, \bibinfo {author} {\bibfnamefont {C.~C.}\ \bibnamefont {Chi}},
  \bibinfo {author} {\bibfnamefont {D.~N.}\ \bibnamefont {Langenberg}},
  \bibinfo {author} {\bibfnamefont {J.~J.}\ \bibnamefont {Chang}}, \bibinfo
  {author} {\bibfnamefont {S.}~\bibnamefont {Jafarey}}, \ and\ \bibinfo
  {author} {\bibfnamefont {D.~J.}\ \bibnamefont {Scalapino}},\ }\Doi
  {10.1103/PhysRevB.14.4854} {\bibfield  {journal} {\bibinfo  {journal} {Phys.
  Rev. B},\ }\textbf {\bibinfo {volume} {14}},\ \bibinfo {pages} {4854}
  (\bibinfo {year} {1976})}\BibitemShut {NoStop}%
\bibitem [{\citenamefont {Martinis}\ \emph {et~al.}(2005)\citenamefont
  {Martinis}, \citenamefont {Cooper}, \citenamefont {McDermott}, \citenamefont
  {Steffen}, \citenamefont {Ansmann}, \citenamefont {Osborn}, \citenamefont
  {Cicak}, \citenamefont {Oh}, \citenamefont {Pappas}, \citenamefont
  {Simmonds},\ and\ \citenamefont {Yu}}]{MartinisTLS}%
  \BibitemOpen
  \bibfield  {author} {\bibinfo {author} {\bibfnamefont {J.~M.}\ \bibnamefont
  {Martinis}}, \bibinfo {author} {\bibfnamefont {K.~B.}\ \bibnamefont
  {Cooper}}, \bibinfo {author} {\bibfnamefont {R.}~\bibnamefont {McDermott}},
  \bibinfo {author} {\bibfnamefont {M.}~\bibnamefont {Steffen}}, \bibinfo
  {author} {\bibfnamefont {M.}~\bibnamefont {Ansmann}}, \bibinfo {author}
  {\bibfnamefont {K.~D.}\ \bibnamefont {Osborn}}, \bibinfo {author}
  {\bibfnamefont {K.}~\bibnamefont {Cicak}}, \bibinfo {author} {\bibfnamefont
  {S.}~\bibnamefont {Oh}}, \bibinfo {author} {\bibfnamefont {D.~P.}\
  \bibnamefont {Pappas}}, \bibinfo {author} {\bibfnamefont {R.~W.}\
  \bibnamefont {Simmonds}}, \ and\ \bibinfo {author} {\bibfnamefont {C.~C.}\
  \bibnamefont {Yu}},\ }\Doi {10.1103/PhysRevLett.95.210503} {\bibfield
  {journal} {\bibinfo  {journal} {Phys. Rev. Lett.},\ }\textbf {\bibinfo
  {volume} {95}},\ \bibinfo {pages} {210503} (\bibinfo {year}
  {2005})}\BibitemShut {NoStop}%
\bibitem [{\citenamefont {McRae}\ \emph {et~al.}(2020)\citenamefont {McRae},
  \citenamefont {Wang}, \citenamefont {Gao}, \citenamefont {Vissers},
  \citenamefont {Brecht}, \citenamefont {Dunsworth}, \citenamefont {Pappas},\
  and\ \citenamefont {Mutus}}]{mcrae}%
  \BibitemOpen
  \bibfield  {author} {\bibinfo {author} {\bibfnamefont {C.~R.~H.}\
  \bibnamefont {McRae}}, \bibinfo {author} {\bibfnamefont {H.}~\bibnamefont
  {Wang}}, \bibinfo {author} {\bibfnamefont {J.}~\bibnamefont {Gao}}, \bibinfo
  {author} {\bibfnamefont {M.~R.}\ \bibnamefont {Vissers}}, \bibinfo {author}
  {\bibfnamefont {T.}~\bibnamefont {Brecht}}, \bibinfo {author} {\bibfnamefont
  {A.}~\bibnamefont {Dunsworth}}, \bibinfo {author} {\bibfnamefont {D.~P.}\
  \bibnamefont {Pappas}}, \ and\ \bibinfo {author} {\bibfnamefont
  {J.}~\bibnamefont {Mutus}},\ }\Doi {10.1063/5.0017378} {\bibfield  {journal}
  {\bibinfo  {journal} {Review of Scientific Instruments},\ }\textbf {\bibinfo
  {volume} {91}},\ \bibinfo {pages} {091101} (\bibinfo {year}
  {2020})}\BibitemShut {NoStop}%
\bibitem [{\citenamefont {De~Visser}\ \emph {et~al.}(2014)\citenamefont
  {De~Visser}, \citenamefont {Baselmans}, \citenamefont {Bueno}, \citenamefont
  {Llombart},\ and\ \citenamefont {Klapwijk}}]{visser2014}%
  \BibitemOpen
  \bibfield  {author} {\bibinfo {author} {\bibfnamefont {P.}~\bibnamefont
  {De~Visser}}, \bibinfo {author} {\bibfnamefont {J.}~\bibnamefont
  {Baselmans}}, \bibinfo {author} {\bibfnamefont {J.}~\bibnamefont {Bueno}},
  \bibinfo {author} {\bibfnamefont {N.}~\bibnamefont {Llombart}}, \ and\
  \bibinfo {author} {\bibfnamefont {T.}~\bibnamefont {Klapwijk}},\ }\href@noop
  {} {\bibfield  {journal} {\bibinfo  {journal} {Nature Communications},\
  }\textbf {\bibinfo {volume} {3}} (\bibinfo {year} {2014})}\BibitemShut
  {NoStop}%
\bibitem [{\citenamefont {de~Visser}\ \emph {et~al.}(2010)\citenamefont
  {de~Visser}, \citenamefont {Withington},\ and\ \citenamefont
  {Goldie}}]{visser2010}%
  \BibitemOpen
  \bibfield  {author} {\bibinfo {author} {\bibfnamefont {P.~J.}\ \bibnamefont
  {de~Visser}}, \bibinfo {author} {\bibfnamefont {S.}~\bibnamefont
  {Withington}}, \ and\ \bibinfo {author} {\bibfnamefont {D.~J.}\ \bibnamefont
  {Goldie}},\ }\Doi {10.1063/1.3517152} {\bibfield  {journal} {\bibinfo
  {journal} {Journal of Applied Physics},\ }\textbf {\bibinfo {volume} {108}},\
  \bibinfo {pages} {114504} (\bibinfo {year} {2010})}\BibitemShut {NoStop}%
\bibitem [{\citenamefont {Thompson}\ \emph {et~al.}(2013)\citenamefont
  {Thompson}, \citenamefont {Withington}, \citenamefont {Goldie},\ and\
  \citenamefont {Thomas}}]{Thompson_2013}%
  \BibitemOpen
  \bibfield  {author} {\bibinfo {author} {\bibfnamefont {S.~E.}\ \bibnamefont
  {Thompson}}, \bibinfo {author} {\bibfnamefont {S.}~\bibnamefont
  {Withington}}, \bibinfo {author} {\bibfnamefont {D.~J.}\ \bibnamefont
  {Goldie}}, \ and\ \bibinfo {author} {\bibfnamefont {C.~N.}\ \bibnamefont
  {Thomas}},\ }\Doi {10.1088/0953-2048/26/9/095009} {\bibfield  {journal}
  {\bibinfo  {journal} {Superconductor Science and Technology},\ }\textbf
  {\bibinfo {volume} {26}},\ \bibinfo {pages} {095009} (\bibinfo {year}
  {2013})}\BibitemShut {NoStop}%
\bibitem [{\citenamefont {Thomas}\ \emph {et~al.}(2015)\citenamefont {Thomas},
  \citenamefont {Withington},\ and\ \citenamefont {Goldie}}]{Thomas_2015}%
  \BibitemOpen
  \bibfield  {author} {\bibinfo {author} {\bibfnamefont {C.~N.}\ \bibnamefont
  {Thomas}}, \bibinfo {author} {\bibfnamefont {S.}~\bibnamefont {Withington}},
  \ and\ \bibinfo {author} {\bibfnamefont {D.~J.}\ \bibnamefont {Goldie}},\
  }\Doi {10.1088/0953-2048/28/4/045012} {\bibfield  {journal} {\bibinfo
  {journal} {Superconductor Science and Technology},\ }\textbf {\bibinfo
  {volume} {28}},\ \bibinfo {pages} {045012} (\bibinfo {year}
  {2015})}\BibitemShut {NoStop}%
\bibitem [{\citenamefont {Cardani}\ \emph {et~al.}(2017)\citenamefont
  {Cardani}, \citenamefont {Casali}, \citenamefont {Colantoni}, \citenamefont
  {Cruciani}, \citenamefont {Bellini}, \citenamefont {Castellano},
  \citenamefont {Cosmelli}, \citenamefont {D'Addabbo}, \citenamefont
  {Di~Domizio}, \citenamefont {Martinez}, \citenamefont {Tomei},\ and\
  \citenamefont {Vignati}}]{Cardani:2017qr}%
  \BibitemOpen
  \bibfield  {author} {\bibinfo {author} {\bibfnamefont {L.}~\bibnamefont
  {Cardani}}, \bibinfo {author} {\bibfnamefont {N.}~\bibnamefont {Casali}},
  \bibinfo {author} {\bibfnamefont {I.}~\bibnamefont {Colantoni}}, \bibinfo
  {author} {\bibfnamefont {A.}~\bibnamefont {Cruciani}}, \bibinfo {author}
  {\bibfnamefont {F.}~\bibnamefont {Bellini}}, \bibinfo {author} {\bibfnamefont
  {M.~G.}\ \bibnamefont {Castellano}}, \bibinfo {author} {\bibfnamefont
  {C.}~\bibnamefont {Cosmelli}}, \bibinfo {author} {\bibfnamefont
  {A.}~\bibnamefont {D'Addabbo}}, \bibinfo {author} {\bibfnamefont
  {S.}~\bibnamefont {Di~Domizio}}, \bibinfo {author} {\bibfnamefont
  {M.}~\bibnamefont {Martinez}}, \bibinfo {author} {\bibfnamefont
  {C.}~\bibnamefont {Tomei}}, \ and\ \bibinfo {author} {\bibfnamefont
  {M.}~\bibnamefont {Vignati}},\ }\Doi {10.1063/1.4974082} {\bibfield
  {journal} {\bibinfo  {journal} {{A}ppl. {P}hys. {L}ett.},\ }\textbf {\bibinfo
  {volume} {110}},\ \bibinfo {pages} {033504} (\bibinfo {year}
  {2017})}\BibitemShut {NoStop}%
\bibitem [{\citenamefont {Swenson}\ \emph {et~al.}(2010)\citenamefont
  {Swenson}, \citenamefont {Cruciani}, \citenamefont {Benoit}, \citenamefont
  {Roesch}, \citenamefont {Yung}, \citenamefont {Bideaud},\ and\ \citenamefont
  {Monfardini}}]{swenson}%
  \BibitemOpen
  \bibfield  {author} {\bibinfo {author} {\bibfnamefont {L.~J.}\ \bibnamefont
  {Swenson}}, \bibinfo {author} {\bibfnamefont {A.}~\bibnamefont {Cruciani}},
  \bibinfo {author} {\bibfnamefont {A.}~\bibnamefont {Benoit}}, \bibinfo
  {author} {\bibfnamefont {M.}~\bibnamefont {Roesch}}, \bibinfo {author}
  {\bibfnamefont {C.~S.}\ \bibnamefont {Yung}}, \bibinfo {author}
  {\bibfnamefont {A.}~\bibnamefont {Bideaud}}, \ and\ \bibinfo {author}
  {\bibfnamefont {A.}~\bibnamefont {Monfardini}},\ }\Doi {10.1063/1.3459142}
  {\bibfield  {journal} {\bibinfo  {journal} {{A}ppl. {P}hys. {L}ett.},\
  }\textbf {\bibinfo {volume} {96}},\ \bibinfo {pages} {263511} (\bibinfo
  {year} {2010})}\BibitemShut {NoStop}%
\bibitem [{\citenamefont {Bourrion}\ \emph {et~al.}(2013)\citenamefont
  {Bourrion}, \citenamefont {Vescovi}, \citenamefont {Catalano}, \citenamefont
  {Calvo}, \citenamefont {D'Addabbo}, \citenamefont {Goupy}, \citenamefont
  {Boudou}, \citenamefont {Macias-Perez},\ and\ \citenamefont
  {Monfardini}}]{Bourrion:2013ifa}%
  \BibitemOpen
  \bibfield  {author} {\bibinfo {author} {\bibfnamefont {O.}~\bibnamefont
  {Bourrion}}, \bibinfo {author} {\bibfnamefont {C.}~\bibnamefont {Vescovi}},
  \bibinfo {author} {\bibfnamefont {A.}~\bibnamefont {Catalano}}, \bibinfo
  {author} {\bibfnamefont {M.}~\bibnamefont {Calvo}}, \bibinfo {author}
  {\bibfnamefont {A.}~\bibnamefont {D'Addabbo}}, \bibinfo {author}
  {\bibfnamefont {J.}~\bibnamefont {Goupy}}, \bibinfo {author} {\bibfnamefont
  {N.}~\bibnamefont {Boudou}}, \bibinfo {author} {\bibfnamefont {J.~F.}\
  \bibnamefont {Macias-Perez}}, \ and\ \bibinfo {author} {\bibfnamefont
  {A.}~\bibnamefont {Monfardini}},\ }\Doi {10.1088/1748-0221/8/12/C12006}
  {\bibfield  {journal} {\bibinfo  {journal} {JINST},\ }\textbf {\bibinfo
  {volume} {8}},\ \bibinfo {pages} {C12006} (\bibinfo {year}
  {2013})}\BibitemShut {NoStop}%
\bibitem [{\citenamefont {Khalil}\ \emph {et~al.}(2012)\citenamefont {Khalil},
  \citenamefont {Stoutimore}, \citenamefont {Wellstood},\ and\ \citenamefont
  {Osborn}}]{khalil}%
  \BibitemOpen
  \bibfield  {author} {\bibinfo {author} {\bibfnamefont {M.~S.}\ \bibnamefont
  {Khalil}}, \bibinfo {author} {\bibfnamefont {M.~J.~A.}\ \bibnamefont
  {Stoutimore}}, \bibinfo {author} {\bibfnamefont {F.~C.}\ \bibnamefont
  {Wellstood}}, \ and\ \bibinfo {author} {\bibfnamefont {K.~D.}\ \bibnamefont
  {Osborn}},\ }\Doi {10.1063/1.3692073} {\bibfield  {journal} {\bibinfo
  {journal} {J. Appl. Phys.},\ }\textbf {\bibinfo {volume} {111}},\ \bibinfo
  {pages} {054510} (\bibinfo {year} {2012})}\BibitemShut {NoStop}%
\bibitem [{\citenamefont {Swenson}\ \emph {et~al.}(2013)\citenamefont
  {Swenson}, \citenamefont {Day}, \citenamefont {Eom}, \citenamefont {LeDuc},
  \citenamefont {Llombart}, \citenamefont {McKenney}, \citenamefont
  {Noroozian},\ and\ \citenamefont {Zmuidzinas}}]{swenson2013}%
  \BibitemOpen
  \bibfield  {author} {\bibinfo {author} {\bibfnamefont {L.}~\bibnamefont
  {Swenson}}, \bibinfo {author} {\bibfnamefont {P.~K.}\ \bibnamefont {Day}},
  \bibinfo {author} {\bibfnamefont {B.~H.}\ \bibnamefont {Eom}}, \bibinfo
  {author} {\bibfnamefont {H.~G.}\ \bibnamefont {LeDuc}}, \bibinfo {author}
  {\bibfnamefont {N.}~\bibnamefont {Llombart}}, \bibinfo {author}
  {\bibfnamefont {C.~M.}\ \bibnamefont {McKenney}}, \bibinfo {author}
  {\bibfnamefont {O.}~\bibnamefont {Noroozian}}, \ and\ \bibinfo {author}
  {\bibfnamefont {J.}~\bibnamefont {Zmuidzinas}},\ }\Doi {10.1063/1.4794808}
  {\bibfield  {journal} {\bibinfo  {journal} {{J}. {A}ppl. {P}hys.},\ }\textbf
  {\bibinfo {volume} {113}},\ \bibinfo {pages} {104501} (\bibinfo {year}
  {2013})}\BibitemShut {NoStop}%
\bibitem [{\citenamefont {Zmuidzinas}(2012)}]{zmu_annrev2012}%
  \BibitemOpen
  \bibfield  {author} {\bibinfo {author} {\bibfnamefont {J.}~\bibnamefont
  {Zmuidzinas}},\ }\Doi {10.1146/annurev-conmatphys-020911-125022} {\bibfield
  {journal} {\bibinfo  {journal} {Annu.Rev.Cond.Mat.Phys.},\ }\textbf {\bibinfo
  {volume} {3}},\ \bibinfo {pages} {169} (\bibinfo {year} {2012})}\BibitemShut
  {NoStop}%
\bibitem [{\citenamefont {Martinez}\ \emph {et~al.}(2019)\citenamefont
  {Martinez}, \citenamefont {Cardani}, \citenamefont {Casali}, \citenamefont
  {Cruciani}, \citenamefont {Pettinari},\ and\ \citenamefont
  {Vignati}}]{martinez2019}%
  \BibitemOpen
  \bibfield  {author} {\bibinfo {author} {\bibfnamefont {M.}~\bibnamefont
  {Martinez}}, \bibinfo {author} {\bibfnamefont {L.}~\bibnamefont {Cardani}},
  \bibinfo {author} {\bibfnamefont {N.}~\bibnamefont {Casali}}, \bibinfo
  {author} {\bibfnamefont {A.}~\bibnamefont {Cruciani}}, \bibinfo {author}
  {\bibfnamefont {G.}~\bibnamefont {Pettinari}}, \ and\ \bibinfo {author}
  {\bibfnamefont {M.}~\bibnamefont {Vignati}},\ }\Doi
  {10.1103/PhysRevApplied.11.064025} {\bibfield  {journal} {\bibinfo  {journal}
  {Phys. Rev. Applied},\ }\textbf {\bibinfo {volume} {11}},\ \bibinfo {pages}
  {064025} (\bibinfo {year} {2019})}\BibitemShut {NoStop}%
\bibitem [{\citenamefont {de~Visser}\ \emph {et~al.}(2014)\citenamefont
  {de~Visser}, \citenamefont {Goldie}, \citenamefont {Diener}, \citenamefont
  {Withington}, \citenamefont {Baselmans},\ and\ \citenamefont
  {Klapwijk}}]{visserwit}%
  \BibitemOpen
  \bibfield  {author} {\bibinfo {author} {\bibfnamefont {P.~J.}\ \bibnamefont
  {de~Visser}}, \bibinfo {author} {\bibfnamefont {D.~J.}\ \bibnamefont
  {Goldie}}, \bibinfo {author} {\bibfnamefont {P.}~\bibnamefont {Diener}},
  \bibinfo {author} {\bibfnamefont {S.}~\bibnamefont {Withington}}, \bibinfo
  {author} {\bibfnamefont {J.~J.~A.}\ \bibnamefont {Baselmans}}, \ and\
  \bibinfo {author} {\bibfnamefont {T.~M.}\ \bibnamefont {Klapwijk}},\ }\Doi
  {10.1103/PhysRevLett.112.047004} {\bibfield  {journal} {\bibinfo  {journal}
  {Phys. Rev. Lett.},\ }\textbf {\bibinfo {volume} {112}},\ \bibinfo {pages}
  {047004} (\bibinfo {year} {2014})}\BibitemShut {NoStop}%
\bibitem [{\citenamefont {de~Visser}\ \emph {et~al.}(2012)\citenamefont
  {de~Visser}, \citenamefont {Baselmans}, \citenamefont {Yates}, \citenamefont
  {Diener}, \citenamefont {Endo},\ and\ \citenamefont
  {Klapwijk}}]{deVisserEta}%
  \BibitemOpen
  \bibfield  {author} {\bibinfo {author} {\bibfnamefont {P.~J.}\ \bibnamefont
  {de~Visser}}, \bibinfo {author} {\bibfnamefont {J.~J.~A.}\ \bibnamefont
  {Baselmans}}, \bibinfo {author} {\bibfnamefont {S.~J.~C.}\ \bibnamefont
  {Yates}}, \bibinfo {author} {\bibfnamefont {P.}~\bibnamefont {Diener}},
  \bibinfo {author} {\bibfnamefont {A.}~\bibnamefont {Endo}}, \ and\ \bibinfo
  {author} {\bibfnamefont {T.~M.}\ \bibnamefont {Klapwijk}},\ }\Doi
  {10.1063/1.4704151} {\bibfield  {journal} {\bibinfo  {journal} {{A}ppl.
  {P}hys. {L}ett.},\ }\textbf {\bibinfo {volume} {100}},\ \bibinfo {pages}
  {162601} (\bibinfo {year} {2012})}\BibitemShut {NoStop}%
\bibitem [{Note1()}]{Note1}%
  \BibitemOpen
  \bibinfo {note} {$\eta _g$ depends on the number of quasiparticles, and thus
  on $P_{qp}$~\protect \cite {Goldie2013}. This dependency can be neglected in
  our first-order model.}\BibitemShut {Stop}%
\bibitem [{\citenamefont {Barends}\ \emph {et~al.}(2008)\citenamefont
  {Barends}, \citenamefont {Baselmans}, \citenamefont {Yates}, \citenamefont
  {Gao}, \citenamefont {Hovenier},\ and\ \citenamefont
  {Klapwijk}}]{BarendsTau}%
  \BibitemOpen
  \bibfield  {author} {\bibinfo {author} {\bibfnamefont {R.}~\bibnamefont
  {Barends}}, \bibinfo {author} {\bibfnamefont {J.~J.~A.}\ \bibnamefont
  {Baselmans}}, \bibinfo {author} {\bibfnamefont {S.~J.~C.}\ \bibnamefont
  {Yates}}, \bibinfo {author} {\bibfnamefont {J.~R.}\ \bibnamefont {Gao}},
  \bibinfo {author} {\bibfnamefont {J.~N.}\ \bibnamefont {Hovenier}}, \ and\
  \bibinfo {author} {\bibfnamefont {T.~M.}\ \bibnamefont {Klapwijk}},\ }\Doi
  {10.1103/PhysRevLett.100.257002} {\bibfield  {journal} {\bibinfo  {journal}
  {Phys. Rev. Lett.},\ }\textbf {\bibinfo {volume} {100}},\ \bibinfo {pages}
  {257002} (\bibinfo {year} {2008})}\BibitemShut {NoStop}%
\bibitem [{Note2()}]{Note2}%
  \BibitemOpen
  \bibinfo {note} {In the calculation of derivative, we approximated $\protect
  \frac {d(\tau _{qp}/Q_{qp})}{dM}=0$ since the dependency of $\tau _{qp}$ and
  $Q_{qp}$ on the number of quasiparticles cancels to a good
  approximation~\protect \cite {deVisserEta}}\BibitemShut {NoStop}%
\bibitem [{\citenamefont {Goldie}\ and\ \citenamefont
  {Withington}(2013)}]{Goldie2013}%
  \BibitemOpen
  \bibfield  {author} {\bibinfo {author} {\bibfnamefont {D.~J.}\ \bibnamefont
  {Goldie}}\ and\ \bibinfo {author} {\bibfnamefont {S.}~\bibnamefont
  {Withington}},\ }\href {http://stacks.iop.org/0953-2048/26/i=1/a=015004}
  {\bibfield  {journal} {\bibinfo  {journal} {Superconductor Science and
  Technology},\ }\textbf {\bibinfo {volume} {26}},\ \bibinfo {pages} {015004}
  (\bibinfo {year} {2013})}\BibitemShut {NoStop}%
\end{thebibliography}%


\begin{thebibliography}{5}%
\makeatletter
\providecommand \@ifxundefined [1]{%
 \@ifx{#1\undefined}
}%
\providecommand \@ifnum [1]{%
 \ifnum #1\expandafter \@firstoftwo
 \else \expandafter \@secondoftwo
 \fi
}%
\providecommand \@ifx [1]{%
 \ifx #1\expandafter \@firstoftwo
 \else \expandafter \@secondoftwo
 \fi
}%
\providecommand \natexlab [1]{#1}%
\providecommand \enquote  [1]{``#1''}%
\providecommand \bibnamefont  [1]{#1}%
\providecommand \bibfnamefont [1]{#1}%
\providecommand \citenamefont [1]{#1}%
\providecommand \href@noop [0]{\@secondoftwo}%
\providecommand \href [0]{\begingroup \@sanitize@url \@href}%
\providecommand \@href[1]{\@@startlink{#1}\@@href}%
\providecommand \@@href[1]{\endgroup#1\@@endlink}%
\providecommand \@sanitize@url [0]{\catcode `\\12\catcode `\$12\catcode
  `\&12\catcode `\#12\catcode `\^12\catcode `\_12\catcode `\%12\relax}%
\providecommand \@@startlink[1]{}%
\providecommand \@@endlink[0]{}%
\providecommand \url  [0]{\begingroup\@sanitize@url \@url }%
\providecommand \@url [1]{\endgroup\@href {#1}{\urlprefix }}%
\providecommand \urlprefix  [0]{URL }%
\providecommand \Eprint [0]{\href }%
\@ifxundefined \urlstyle {%
  \providecommand \doi  [0]{\begingroup \@sanitize@url \@doi}%
  \providecommand \@doi [1]{\endgroup \@@startlink {\doibase
  #1}doi:\discretionary {}{}{}#1\@@endlink }%
}{%
  \providecommand \doi  [0]{doi:\discretionary{}{}{}\begingroup
  \urlstyle{rm}\Url }%
}%
\providecommand \doibase [0]{http://dx.doi.org/}%
\providecommand \Doi [0]{\begingroup \@sanitize@url \@Doi }%
\providecommand \@Doi  [1]{\endgroup\@@startlink{\doibase#1}\@@Doi}%
\providecommand \@@Doi [1]{#1\@@endlink}%
\providecommand \selectlanguage [0]{\@gobble}%
\providecommand \bibinfo  [0]{\@secondoftwo}%
\providecommand \bibfield  [0]{\@secondoftwo}%
\providecommand \translation [1]{[#1]}%
\providecommand \BibitemOpen [0]{}%
\providecommand \bibitemStop [0]{}%
\providecommand \bibitemNoStop [0]{.\EOS\space}%
\providecommand \EOS [0]{\spacefactor3000\relax}%
\providecommand \BibitemShut  [1]{\csname bibitem#1\endcsname}%
\bibitem [{\citenamefont {Colantoni}\ \emph {et~al.}(2016)\citenamefont
  {Colantoni}, \citenamefont {Bellini}, \citenamefont {Cardani}, \citenamefont
  {Casali}, \citenamefont {Castellano}, \citenamefont {Coppolecchia},
  \citenamefont {Cosmelli}, \citenamefont {Cruciani}, \citenamefont
  {D'Addabbo}, \citenamefont {Di~Domizio} \emph {et~al.}}]{Colantoni2016}%
  \BibitemOpen
  \bibfield  {author} {\bibinfo {author} {\bibfnamefont {I.}~\bibnamefont
  {Colantoni}}, \bibinfo {author} {\bibfnamefont {F.}~\bibnamefont {Bellini}},
  \bibinfo {author} {\bibfnamefont {L.}~\bibnamefont {Cardani}}, \bibinfo
  {author} {\bibfnamefont {N.}~\bibnamefont {Casali}}, \bibinfo {author}
  {\bibfnamefont {M.~G.}\ \bibnamefont {Castellano}}, \bibinfo {author}
  {\bibfnamefont {A.}~\bibnamefont {Coppolecchia}}, \bibinfo {author}
  {\bibfnamefont {C.}~\bibnamefont {Cosmelli}}, \bibinfo {author}
  {\bibfnamefont {A.}~\bibnamefont {Cruciani}}, \bibinfo {author}
  {\bibfnamefont {A.}~\bibnamefont {D'Addabbo}}, \bibinfo {author}
  {\bibfnamefont {S.}~\bibnamefont {Di~Domizio}},  \emph {et~al.},\ }\Doi
  {10.1007/s10909-015-1452-1} {\bibfield  {journal} {\bibinfo  {journal}
  {Journal of Low Temperature Physics},\ \bibinfo {pages} {1}} (\bibinfo {year}
  {2016})},\ ISSN \bibinfo {issn} {1573-7357}\BibitemShut {NoStop}%
\bibitem [{\citenamefont {Bourrion}\ \emph {et~al.}(2013)\citenamefont
  {Bourrion}, \citenamefont {Vescovi}, \citenamefont {Catalano}, \citenamefont
  {Calvo}, \citenamefont {D'Addabbo}, \citenamefont {Goupy}, \citenamefont
  {Boudou}, \citenamefont {Macias-Perez},\ and\ \citenamefont
  {Monfardini}}]{Bourrion:2013ifa}%
  \BibitemOpen
  \bibfield  {author} {\bibinfo {author} {\bibfnamefont {O.}~\bibnamefont
  {Bourrion}}, \bibinfo {author} {\bibfnamefont {C.}~\bibnamefont {Vescovi}},
  \bibinfo {author} {\bibfnamefont {A.}~\bibnamefont {Catalano}}, \bibinfo
  {author} {\bibfnamefont {M.}~\bibnamefont {Calvo}}, \bibinfo {author}
  {\bibfnamefont {A.}~\bibnamefont {D'Addabbo}}, \bibinfo {author}
  {\bibfnamefont {J.}~\bibnamefont {Goupy}}, \bibinfo {author} {\bibfnamefont
  {N.}~\bibnamefont {Boudou}}, \bibinfo {author} {\bibfnamefont {J.~F.}\
  \bibnamefont {Macias-Perez}}, \ and\ \bibinfo {author} {\bibfnamefont
  {A.}~\bibnamefont {Monfardini}},\ }\Doi {10.1088/1748-0221/8/12/C12006}
  {\bibfield  {journal} {\bibinfo  {journal} {JINST},\ }\textbf {\bibinfo
  {volume} {8}},\ \bibinfo {pages} {C12006} (\bibinfo {year}
  {2013})}\BibitemShut {NoStop}%
\bibitem [{cit()}]{citlf3}%
  \BibitemOpen
  \href {https://www.cosmicmicrowavetechnology.com/citlf3} {\enquote {\bibinfo
  {title} {{C}osmic {M}icrowave {T}echnology, {I}nc -- {CITLF3}},}\
  }\BibitemShut {NoStop}%
\bibitem [{\citenamefont {Swenson}\ \emph {et~al.}(2013)\citenamefont
  {Swenson}, \citenamefont {Day}, \citenamefont {Eom}, \citenamefont {LeDuc},
  \citenamefont {Llombart}, \citenamefont {McKenney}, \citenamefont
  {Noroozian},\ and\ \citenamefont {Zmuidzinas}}]{swenson2013}%
  \BibitemOpen
  \bibfield  {author} {\bibinfo {author} {\bibfnamefont {L.}~\bibnamefont
  {Swenson}}, \bibinfo {author} {\bibfnamefont {P.~K.}\ \bibnamefont {Day}},
  \bibinfo {author} {\bibfnamefont {B.~H.}\ \bibnamefont {Eom}}, \bibinfo
  {author} {\bibfnamefont {H.~G.}\ \bibnamefont {LeDuc}}, \bibinfo {author}
  {\bibfnamefont {N.}~\bibnamefont {Llombart}}, \bibinfo {author}
  {\bibfnamefont {C.~M.}\ \bibnamefont {McKenney}}, \bibinfo {author}
  {\bibfnamefont {O.}~\bibnamefont {Noroozian}}, \ and\ \bibinfo {author}
  {\bibfnamefont {J.}~\bibnamefont {Zmuidzinas}},\ }\Doi {10.1063/1.4794808}
  {\bibfield  {journal} {\bibinfo  {journal} {{J}. {A}ppl. {P}hys.},\ }\textbf
  {\bibinfo {volume} {113}},\ \bibinfo {pages} {104501} (\bibinfo {year}
  {2013})}\BibitemShut {NoStop}%
\bibitem [{\citenamefont {Khalil}\ \emph {et~al.}(2012)\citenamefont {Khalil},
  \citenamefont {Stoutimore}, \citenamefont {Wellstood},\ and\ \citenamefont
  {Osborn}}]{khalil}%
  \BibitemOpen
  \bibfield  {author} {\bibinfo {author} {\bibfnamefont {M.~S.}\ \bibnamefont
  {Khalil}}, \bibinfo {author} {\bibfnamefont {M.~J.~A.}\ \bibnamefont
  {Stoutimore}}, \bibinfo {author} {\bibfnamefont {F.~C.}\ \bibnamefont
  {Wellstood}}, \ and\ \bibinfo {author} {\bibfnamefont {K.~D.}\ \bibnamefont
  {Osborn}},\ }\Doi {10.1063/1.3692073} {\bibfield  {journal} {\bibinfo
  {journal} {J. Appl. Phys.},\ }\textbf {\bibinfo {volume} {111}},\ \bibinfo
  {pages} {054510} (\bibinfo {year} {2012})}\BibitemShut {NoStop}%
\end{thebibliography}%

\end{document}


\title{Non-linearity in the system of quasiparticles of a superconducting resonator: Supplementary material}
\author {M.~Vignati}
\email[Corresponding author: ]{marco.vignati@roma1.infn.it}
\affiliation{Sapienza Universit\`a di Roma -- Dipartimento di Fisica, I-00185, Roma, Italy}
\affiliation{Istituto Nazionale di Fisica Nucleare -- Sezione di Roma, I-00185, Roma, Italy}
\author{C.~Bellenghi}
\altaffiliation{Now at Technische Universit\"at M\"unchen, Physik-Department,  D-85748, Garching, Germany}
\affiliation{Sapienza Universit\`a di Roma -- Dipartimento di Fisica, I-00185, Roma, Italy}
\author{L.~Cardani}
\author{N.~Casali}
\affiliation{Istituto Nazionale di Fisica Nucleare -- Sezione di Roma, I-00185, Roma, Italy}
\author{I.~Colantoni}
\affiliation{Consiglio Nazionale delle Ricerche -- Istituto di Nanotecnologia, I-00185, Roma, Italy}
\affiliation{Istituto Nazionale di Fisica Nucleare -- Sezione di Roma, I-00185, Roma, Italy}
\author{A.~Cruciani}
\affiliation{Istituto Nazionale di Fisica Nucleare -- Sezione di Roma, I-00185, Roma, Italy}

\date{\today}
\maketitle

\section{Experimental setup}
\begin{figure}[h]
\begin{center}
\includegraphics[width=0.35\linewidth]{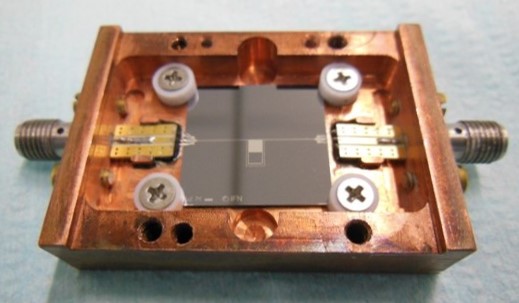}
\includegraphics[width=0.52\linewidth]{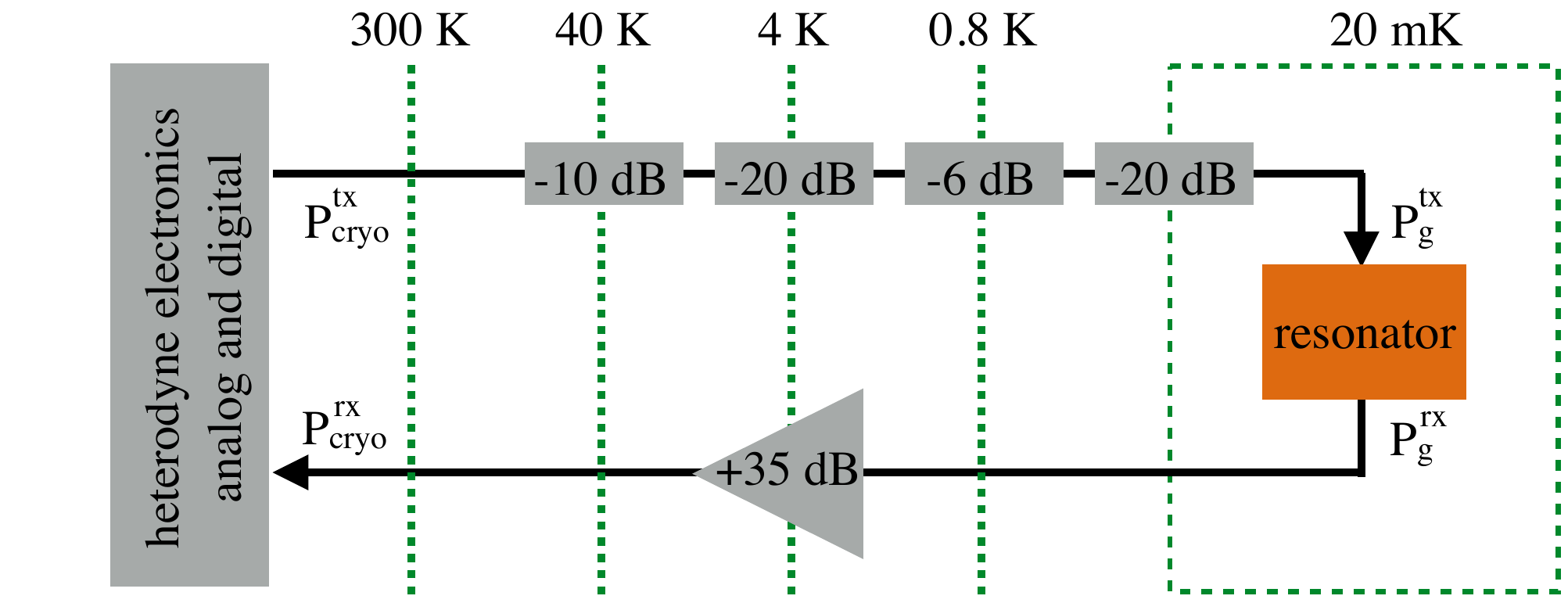}
\caption{
{\bf Left:} photo of the aluminum resonator on the silicon substrate mounted in the copper holder;
{\bf right:} readout scheme of the resonator placed in the dilution refrigerator. 
}
\label{fig:chip}
\end{center}
\end{figure}

The device reported in the present work was fabricated on a high-quality (FZ method) intrinsic silicon (100) substrate, 
with high resistivity and double side polished. 
The resonator and the CPW were patterned by electron beam lithography on a single 60~nm thick aluminum film, deposited using electron-gun evaporation (see Ref.~\cite{Colantoni2016} for more details). The chip was mounted in a copper holder, fixed using PTFE supports, connected to SMA read-out using wedge bonding (see Fig.~\ref{fig:chip} left), and then installed  at the coldest point of a $^{3}$He/$^{4}$He dilution refrigerator. 

The readout scheme is shown in Fig.~\ref{fig:chip} (right).
The microwave generated at room temperature by the electronics~\cite{Bourrion:2013ifa} enters the cryostat with power $P_{\rm cryo}^{\rm tx}$, is reduced by 56~dB  down to $P_{\rm g}^{\rm tx}$ with  attenuators placed at the different temperature stages of the cryostat  and is sent to the CPW on the chip. 
The microwave at the output of the chip, with power $P_{\rm g}^{\rm rx}$,  is driven to a CITLF-3 amplifier~\cite{citlf3} placed at the 4~K temperature stage and featuring an amplification of 35 dB at 2.5~GHz. Past the amplifier the signal  exits the cryostat with power $P_{\rm cryo}^{\rm rx}$ and is sent back to the electronics.

When the resonator is driven at off-resonance frequencies, in principle $P_{\rm g}^{\rm tx}=P_{\rm g}^{\rm rx}$. However by measuring $P_{\rm cryo}^{\rm tx}$ and $P_{\rm cryo}^{\rm rx}$ and by tracing back the powers at the resonator including the attenuation of the cables, 
we find $P_{\rm g}^{\rm rx}/P_{\rm g}^{\rm tx}=- 3~\rm{dB}$, likely because of the uncertainty in the calibration of the components of the transmission line and because of impedance mismatches at the interface with the CPW.  The values of $P_g$ quoted in the main text are the mean values between $P_{\rm g}^{\rm tx}$ and $P_{\rm g}^{\rm rx}$, which are not used to derive the results but are reported for comparison with other devices.
\vspace{1cm}

\newpage
\section{Analysis of $S_{21}$ data}
\noindent
The values of $Q/Q_c$ and of $y$ are extracted directly from the real and imaginary parts of the $S_{21}$ data.  
From Eq.~4 we have:
\begin{eqnarray} 
\Re{(1-S_{21})} &=& \frac{Q}{Q_c}\frac{1}{1+4y^2}\\
\Im{(1-S_{21})} &=& -\frac{Q}{Q_c}\frac{2y}{1+4y^2}
\end{eqnarray}
so that:
\begin{eqnarray}
y &=& - \frac{1}{2}\frac{\Im{(1-S_{21})}}{\Re{(1-S_{21})}}\\
v = \frac{Q}{Q_c} &=& \frac{ \Re^2{(1-S_{21})} + \Im^2{(1-S_{21})} } {\Re{(1-S_{21})}} 
\end{eqnarray}
The advantage of this method is that any dependency of the quality factor on the absorbed power is directly
estimated point by point from the $S_{21}$ data without any interpolation.  

\begin{figure}[t]
\begin{center}
\begin{overpic}[width=0.33\linewidth]{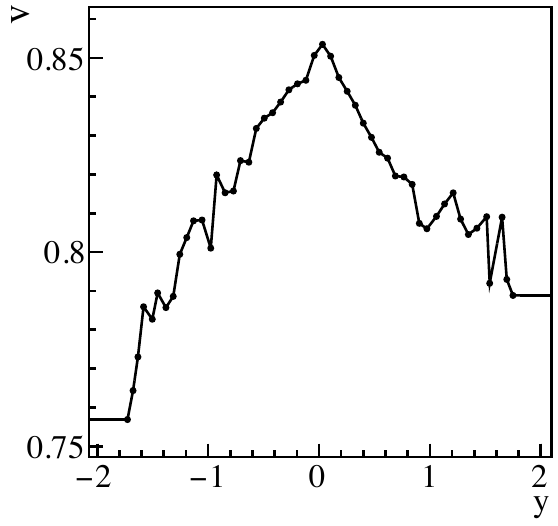}
 \put (75,85) {{\textbf a)}}
\end{overpic}
\begin{overpic}[width=0.33\linewidth]{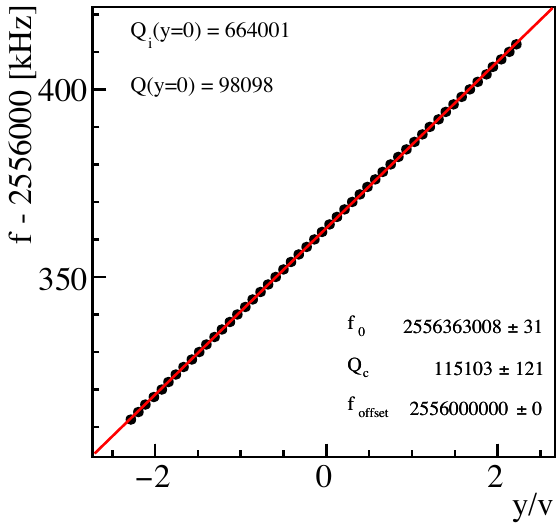}
 \put (75,85) {{\textbf b)}}
\end{overpic}
\begin{overpic}[width=0.33\linewidth]{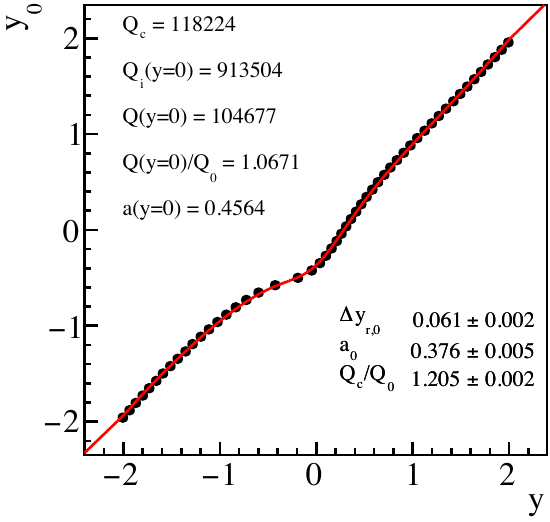}
 \put (75,85) {{\textbf c)}}
\end{overpic}
\begin{overpic}[width=0.33\linewidth]{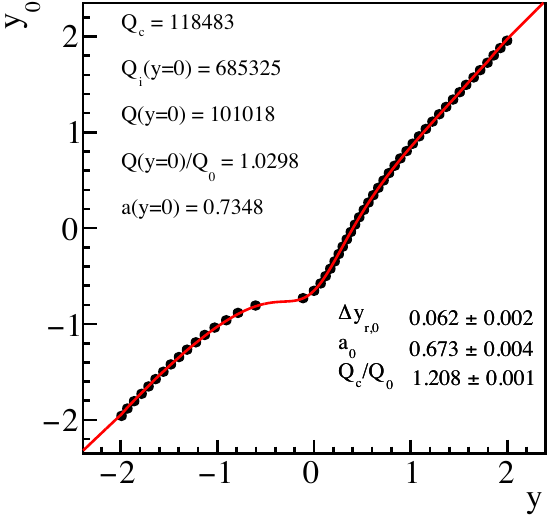}
 \put (75,85) {{\textbf d)}}
\end{overpic}
\caption{
{\textbf a)} $v$ vs $y$ values for the data at low power;
{\textbf b)} generator frequency $f_g$ as a function of $y$ for data at low power. The line is a fit for   $f_{r,0}$ and $Q_c$ with Eq.~\ref{eq:scanfit_init},  the parameters $Q_i(y=0)$ and Q(y=0) are calculated as described in the text;
{\textbf c)} fit of data at medium power with the 3-parameter model (Eq.~\ref{eq:3param}); 
{\textbf d)} fit of data at high power with the 3-parameter model. 
}
\label{fig:resonancefit}
\end{center}
\end{figure}

The characteristics of the resonator $Q_c,Q_0$ and $f_{r,0}$ are evaluated from the low power data.
By setting $a=0$ in 
Eq.~10, 
we have that:
\begin{equation}\label{eq:scanfit_init}
y_0 = \frac{y}{z} \rightarrow f_g = f_{r,0}\left(1  +  \frac{y}{v}\frac{1}{Q_c}\right) 	
\end{equation}
The above equation is used to extract $f_{r,0}$ and $Q_c$ with a linear fit of $f_g$ vs $y/v$ (Fig.~\ref{fig:resonancefit} a and b).
The value of $Q_0$ is then calculated as:
\begin{equation}
Q_0 = v(y=0)Q_c
\end{equation} 
and from the definition of the quality factor ($Q^{-1} = Q_c^{-1} + Q_i^{-1}$), one can  derive the internal quality factor.
It has to be stressed that $Q_0$ has to be considered
only as reference value, i.e. the value we obtain when biasing the circuit on-resonance with $P_g = 1.6$~fW. Another choice of readout power would have implied a different value of $Q_0$, given the dependency of  the internal quality factor on the absorbed power.

Once $f_{r,0}, Q_0$ and $Q_c$ are fixed to the low-power values, the values of $y_0$ and $z=v Q_c / Q_0$ are calculated for the data points at all powers and then Eq.~10 is used to fit the data. Since the equation  has a 3rd degree dependency on $y$, the solution features discontinuities when the resonator is in the bifurcation regime (see Ref.~\cite{swenson2013}). Therefore we chose to invert the equation and fit $y_0$ as a function of $y$ to ease the minimization procedure:
\begin{equation}\label{eq:1parm}
y_0 = \frac{y}{z} -  \frac{a_0 z^2}{1+4y^2}
\end{equation}
As it can be seen from Fig.~2 (right) the fits do not perfectly match the data. One of the reasons could be that the impedance of the resonator and of the coplanar-wave-guide, which is also made of aluminum, may change with power and thus $Q_c$ could slightly vary.
Also, because of the presence of TLS, $f_{r,0}$ could slightly increase with power. In order to take into account these effects, the above equation can be therefore modified as it follows:
\begin{equation}\label{eq:3param}
y_0 = \Delta y_{r,0} + \frac{y}{v}\frac{Q_0}{Q_c} -  \left(\frac{vQ_c}{Q_0}\right)^2\frac{a_0}{1+4y^2}
\end{equation}
where $\Delta y_{r,0}$ accounts for the resonant frequency shift  and the free parameters are $\Delta y_{r,0},Q_c/Q_0$ and $a_0$.  The fits with this 3-parameter model better reproduce the behavior of the ($y_0,y$) data (see~Fig.~\ref{fig:resonancefit} c and d),  with $a_0$ values around 6\% lower than  the results from the fits with the model quoted in the main text (Fig. 2). We consider the discrepancy between models as systematic error on the estimation of $a_0$.

In order to reduce the systematic error, however, one should  take into account of the $Q_c$ correlation with the procedure applied to deconvolve  the transmission line effects~\cite{khalil}. Extra calibrations of the transmission-line with power should be performed, but they were outside the scope of this work since the results presented are considered robust. 

\section{Estimation of $\delta x$ and $\delta Q^{-1}/\delta x$}
The pulses are measured as phase ($\delta\phi$) and amplitude ($\delta A$) variations of $S_{21}$ with respect to the center of the resonance ``circle''. 
It has to be noticed that, given the variation of $Q$ with $y$, the $S_{21}$ data do not lye on a perfect geometrical circle. However, given that this variation amounts at maximum  to 5\% at medium at high powers (see Fig.2 left), we still fit the data with a circle function to evaluate its center and radius.

From Eq.~4 the variations are:
\begin{eqnarray}
\delta \phi &=& 4 \frac{Q\delta x}{1+4y^2}\\
\delta A &=& 2 \frac{Q \delta Q^{-1}}{1+4y^2}
\end{eqnarray}
where $\delta A$ is normalized to the radius of the circle. Therefore the quantities $\delta x/\delta x_{qp}$ and $\delta Q^{-1}/\delta x$ can be calculated as:
\begin{eqnarray}
\frac{\delta x}{\delta x_{qp}} &=& \frac{z_H}{z} \frac{(1+4y^2)}{(1+4y_H^2)}\frac{\delta \phi}{\delta \phi_H} \\
\frac{\delta Q^{-1}}{dx} &=& 2\frac{\delta A}{\delta \phi}
\end{eqnarray}
were $\delta \phi_H$ is the phase variation at the maximum detuning $y_H$, where feedback effects are suppressed (see Eq. 16).

\begin{figure}[tb]
\begin{center}
\includegraphics[width=0.8\linewidth]{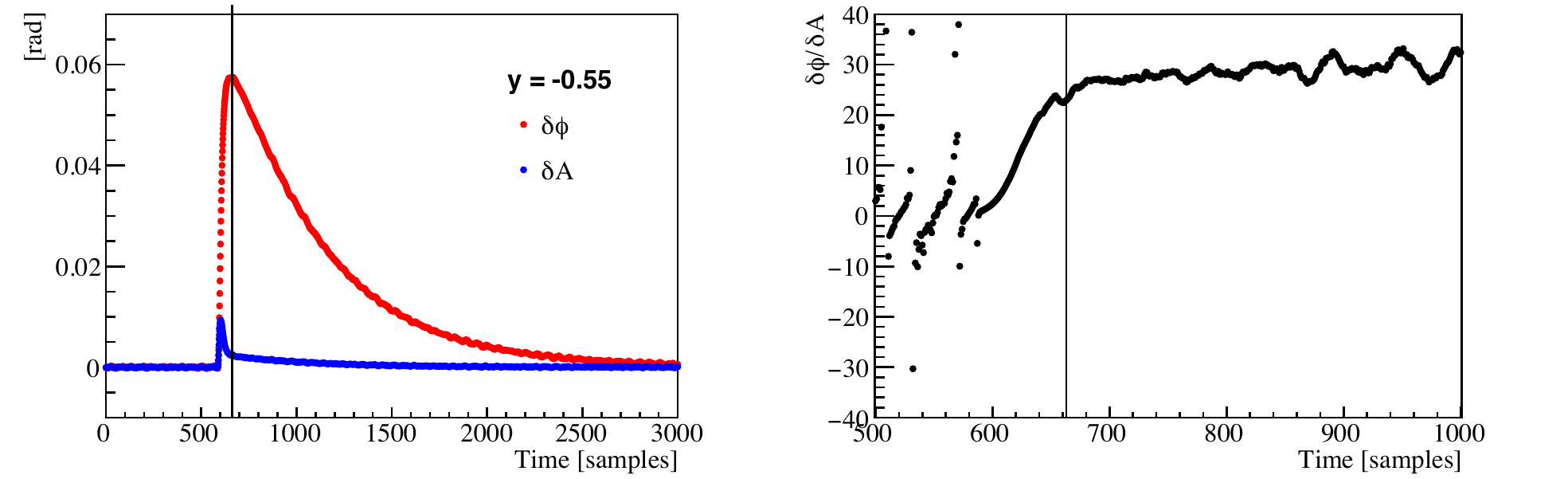}
\includegraphics[width=0.8\linewidth]{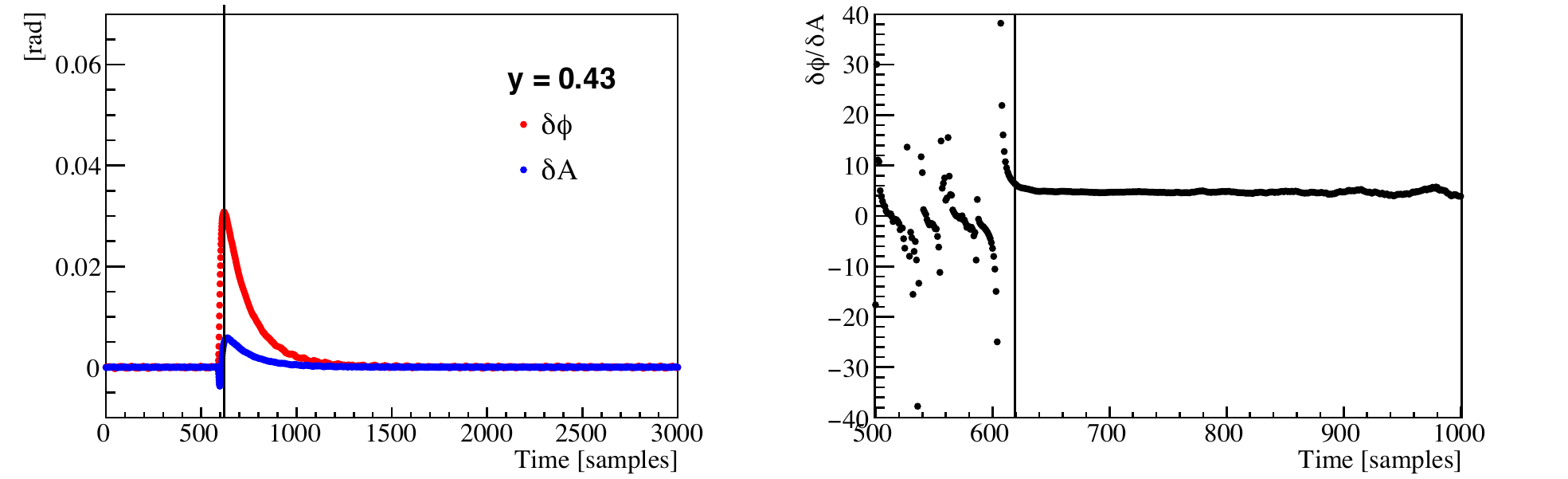}
\caption{
{\bf Left}: Average $\delta\phi$ (red) and $\delta A$ (blue) waveforms acquired with $2~\mu$s sampling time. Data taken at high power for $y=-0.55$ (top) and $y=0.43$ (bottom); 
{\bf right}: Ratio between waveforms  ($\delta\phi/\delta A$) zoomed around the pulse region. The lines indicate the position of the maximum of the $\delta\phi$ pulses.
The ratio in correspondence of the raising edge is not constant because of the different shape with detuning. The trailing edge has the same shape and the average is used as estimate of $\delta\phi/\delta A$.
}
\label{fig:betafrompulses}
\end{center}
\end{figure}

The transfer function of the resonator, which at $y=0$ and at low-power is a single-pole low-pass filter with cutoff equal to $1/(2\pi\tau_r)$,
becomes asymmetric between positive and negative signal frequencies when  $y\neq0$ at low power and at any $y$ in the non-linear regime.
This implies that the $\delta$-response of the resonator is no longer real, but contains an imaginary part.
This effect mixes  the $\delta\phi$ and $\delta A$ responses at  frequencies around and above $1/2\pi\tau_r$.
Since  $\delta\phi\gg\delta A$ the mixing affects evidently only $\delta A$, and in particular the raising edge of the pulse which is of the order of $\tau_r$. 
Figure~\ref{fig:betafrompulses} (left) shows the $\delta\phi$ and $\delta A$ waveforms at high power and for relatively high detunings. As it can be seen the $\delta A$ pulse exhibit an overshoot ($y<0$) or an undershoot ($y>0$)
due to this ``asymmetric cutoff''. Therefore while  $\delta \phi$ is estimated directly from the pulse amplitude, the value of $\delta Q^{-1}/dx$ is evaluated from the ratio of the trailing edges of the pulses, i.e. at signal frequencies well below the cutoff of the resonator since $\tau_{rel} \gg \tau_r$ (Fig.~\ref{fig:betafrompulses} right).  

\bibliographystyle{apsrev4-1}
\bibliography{../../calder}